\documentclass[fleqn,usenatbib]{mnras}
\pdfoutput=1
\usepackage{newtxtext,newtxmath}

\usepackage[T1]{fontenc}
\usepackage{ae,aecompl}
\usepackage{physics}

\usepackage{graphicx}	
\usepackage{epstopdf}
\usepackage{psfig} 
\usepackage{amsmath}	
\usepackage{amssymb}	
\usepackage[T1]{fontenc}
\usepackage{aecompl}
\usepackage{multirow}
\usepackage{subcaption}	
\usepackage{enumitem}
\title[Analysing velocity structure]{A method to analyse velocity structure}

\author[Becky Arnold et al.]{
Becky Arnold,\thanks{E-mail: rjarnold1@sheffield.ac.uk}
and Simon P. Goodwin
\\
Department of Physics and Astronomy, University of Sheffield, Sheffield S3 7RH, UK\\
}

\date{Accepted XXX. Received YYY; in original form ZZZ}

\pubyear{2018}

\begin{document}
\label{firstpage}
\pagerange{\pageref{firstpage}--\pageref{lastpage}}
\maketitle

\begin{abstract}

  We present a new method of analysing and 
  quantifying velocity structure in star forming regions suitable for the rapidly 
  increasing quantity and quality of stellar position--velocity data.
  The method can be applied to data in any number of dimensions, does not
  require the centre or characteristic size (e.g. radius) of the region to be determined, and
  can be applied to regions with any underlying density and velocity structure.
    We test the method on a variety of example datasets and show it is robust with 
    realistic observational uncertainties and selection effects.  This method identifies
    velocity structures/scales in a region, and allows a direct comparison to be made between regions.

\end{abstract}




\section{Introduction}

Star forming regions are an important part of our
understanding of the universe. Their formation and evolution
has important implications for our grasp of planet formation,
star formation, and stellar evolution.
  
In an effort to understand these regions and their evolution
several methods have been developed for quantifying aspects of their
spatial structure. For example the Q parameter \citep{Cartwright04}
describes the degree of spatial
substructure in a region which aids investigations into how substructured
regions evolve. The $\Lambda$ \citep{Alison09a}
and $\Sigma$ \citep{Maschberger11} parameters evaluate the degree of
mass segregation in a region which has significant implications
for our understanding of how massive stars form, how clusters form,
and how clusters evolve.
  
Such methods of quantifying spatial structure have proved valuable
and are well used, but there are not corresponding widely used methods for
quantifying velocity structure. In absence of such methods several approaches
have been used. The most basic approach is
to look at the raw velocity data, often in the form of arrows overplotted on
physical space, e.g. \citet{Galli13}, \citet{Kounkel18}.
This is taken further in \citet{Wright16} and \citet{Wright18} which colour code arrows
according to their direction. This approach can be helpful for
getting a sense of a region's velocity structure, but does not provide an
objective output that quantifies it. As a result interpretation based on
this alone is often subjective.
\citet{Wright16} and \citet{Wright18} also perform spatial correlation tests
to confirm the presence of kinematic substructure in their datasets, but these
tests can say little about the distribution of that substructure.

\citet{Alfaro16} presents a minimum spanning tree based method of quantifying
kinematic substructure. This method also provides graphical indications of how
this substructure is distributed. However, it is primarily designed for
(and solely applied to) radial velocity datasets.

  Another tool that has been used to study velocity structure is the
  PPV (position-position-velocity) diagram which plots stellar positions on two
  axes and one velocity component on a third, e.g \citet{DaRio17}. 
  Efforts to include extra velocity components using, for example, colour-coding
  or different sized data points generally make the diagram far too complex to
  reasonably interpret. It is also difficult to display multidimensional errorbars.
  This limits the usefulness of PPV diagrams when the
  third spatial component and/or additional velocity components are measured.

The lack of objective, quantitative tools for studying kinematic substructure can
in part be attributed to
a previous absence of significant quantities of high-quality velocity data.
However, the next few years will see a revolution in kinematic data for Galactic astrophysics
due to \textit{Gaia}, large multi-object spectroscopy radial velocity surveys, and 
longer time-baseline proper motion studies. With more and more position-velocity 
data becoming available we need tools with which to analyse and interpret it. 

In this paper we introduce a new method for analysing velocity structure, 
borrowing from the concept of variograms (a tool used in geology), which are based on
principles introduced in \citet{Krige51}, and formalised in \citet{Matheron63}.
Here the method is discussed in the context
of analysing velocity structure in star forming regions, but the method
is extremely general and can be applied to regions of any size and
morphology. This makes it well suited for objectivley comparing
very different regions. The method can also be applied to datasets in any number of dimensions
without additional difficulty and it does not demand that the position
and velocity data is in the same number of dimensions. High dimensional datasets are often hard
to visualise and apprehend, so this method aids the interpretation of such datasets
(e.g. as provided by $Gaia$). Its quantative nature also makes it well suited to
objectively analysing the degree of kinematic substructure in a region.
Examples of datasets that this method could be applied to include \citet{Wright16}, \citet{Wright18},
\citet{Gagne18}, \citet{Franciosini18} and \citet{Kuhn18_prep}.

A program called the Velocity Structure Analysis Tool, \textsc{\small{}vsat},  which
runs this method is available at
\hyperlink{https://github.com/r-j-arnold/VSAT}{https://github.com/r-j-arnold/VSAT}.

\section{The VSAT method} \label{Method}

We outline the method below before applying it to a variety of test datasets.

In brief, for every possible pair of stars the distance between 
them ($\Delta r$) is calculated along with the pairs velocity
difference ($\Delta v$). Pairs are then sorted into $\Delta r$ bins.
In each bin the mean $\Delta v$ of the pairs it contains is calculated.
These mean $\Delta v$ values are then plotted against their corresponding
$\Delta r$ values. The values and shape of this distribution can be used to 
understand the velocity structure of the region and can be directly compared 
with those produced for any other region (i.e. they are in informative physical 
values of km s$^{-1}$ and pc).

The method is applied twice, each using a different definition of velocity difference, $\Delta v$,
which highlight different aspects of a region's velocity structure. 
The first definition is referred to as the magnitude definition, $\Delta v_{\rm M}$. 
If star $i$ has velocity vector $\boldsymbol{v}_i$ and star $j$ has velocity vector $\boldsymbol{v}_j$ then
$\Delta v_{\rm M}$ is the magnitude of their difference,  $\mid \boldsymbol{v}_i - \boldsymbol{v}_j \mid $.
We stress that $\Delta v_{\rm M}$  is 
the \textit{magnitude of the difference} of the star's velocities, and {\em not the difference of the magnitudes.}
The equation to calculate $\Delta v_{\rm M}$  (assuming two dimensions for simplicity) is:
\begin{equation} \label{dv_orig}
  \Delta v_{i\!j \rm{M}} ~=~ \sqrt{(v_{x\!i} - v_{x\!j})^{2} ~+~ (v_{y\!i} - v_{y\!j})^{2}}.
\end{equation}
As $\Delta v_{\rm M}$ is a magnitude it is always positive.

The other definition of $\Delta v$ is referred to as the directional definition, $\Delta v_{\rm D}$.
It is the rate at which the distance between the stars, $\Delta r$, is changing, i.e. it is how fast the stars are moving towards/away from one another.
This value is positive if $\Delta r$ is increasing (stars are
moving away from each other), negative if $\Delta r$ is decreasing (they are moving towards each other), and zero if they 
are not moving relative to each other. As such this could be considered a measure of velocity divergence.
In two dimensions the equation to calculate $\Delta v_{\rm D}$ is:

\begin{equation}
  \Delta v_{i\!j \rm{D}} ~=~ \frac{(x_{i} - x_{j})(v_{x\!i} - v_{x\!j}) ~+~ (y_{i} - y_{j})(v_{y\!i} - v_{y\!j})}{\Delta r_{i\!j}}.
\end{equation}

This definition is particularly useful for investigating if a region (or structures within a region)
are expanding or collapsing.

The method makes no assumptions about the underlying distribution of the star's 
positions or velocities and does not require the region's radius or centre to 
be defined. We show in section \ref{apply_obs_errs} that it is relatively insensitive to even quite 
large observational uncertainties and biases, and works reasonably even when $N$ is small ($<100$).  

Throughout we will assume that the data we are dealing with is 2D velocities 
(proper motion) and 2D positions: i.e. what would be provided by \textit{Gaia} 
with good precision (and what is also simple to present in a figure). It is 
trivial to extend the method to full 6D information from simulations, or to 
add radial velocities (with a different uncertainty), or indeed any combination of 
spatial and velocity dimensions.
 
A full step-by-step explanation of the method now follows.

\begin{enumerate}[label=\textbf{\arabic{enumi}}, leftmargin=*]

\item \textbf{Calculate $\Delta r$ and $\Delta v$
for every possible pair of stars.} 

For any pair of stars $i$ and $j$
their separation $\Delta r_{i\!j}$ is (in 2D):
\begin{equation} \label{dr_orig}
    \Delta r_{i\!j} ~=~ \sqrt{(x_{i} - x_{j})^{2} ~+~ (y_{i} - y_{j})^{2}}.
\end{equation}

Calculate $\Delta v$ using either the magnitude or directional definition as desired.
Note that as all measures are relative the frame of reference is irrelevant 
(i.e. there is no need to shift into a centre-of-mass or -velocity frame).

\item \textbf{Calculate errors on $\Delta v$.} 
  
  If there are observational errors 
  propagate them to
  calculate $\sigma_{\Delta v_{i\!j}}$, the error on each $\Delta v_{i\!j}$.
  A measurement $\Delta v_{i\!j}$
  has weight
  \begin{equation}
    w_{i\!j} ~=~ \frac{1}{\sigma_{\Delta v_{i\!j}}^{2}}.
  \end{equation}
  Observational errors on stellar positions are typically {\em much} smaller than on their
  velocities, so they are neglected in this paper.

\item \textbf{Sort the pairs into $\Delta r$ bins.}  
  
  Each bin should 
  contain a significant ($>>30$) number of pairings, but because the 
  number of pairings scales as $N^2$ (where $N$ is the number of stars 
  in the dataset) even fairly low $N$ will result in a relatively large 
  number of pairings.  As long as the number of pairings in each bin is 
  large the bin widths have very little impact on the results (in the 
  examples shown later we use bins of width 0.1~pc and most bins contain $>$ 1000 pairs).

\item \label{obs_err_step} \textbf{For each $\Delta r$
  bin calculate the mean $\Delta v$ of the pairs
  it contains, $\Delta v(\Delta r)$}. 
  
  This gives the mean velocity difference
  of stars separated by a given $\Delta r$.  
  
  In the case that there are observational errors use the weighted mean for this step. 
  The uncertainty on this mean due to observational errors is:
  \begin{equation}
    \sigma_{\rm obs} ~=~ \sqrt{\frac{1}{\sum w_{i\!j}}}
  \end{equation}
  Where the sum is over the pairs of stars $i\!j$ in the bin.    
    
\item \textbf{Calculate errors due to stochasticity.} 

  The value of
  $\Delta v(\Delta r)$ calculated for each bin obviously depends on the
  precise positions and velocities of the stars.  
  
  However, even in `perfect' data there is a stochastic error 
  due to the sampling of an underlying distribution with $N$ points.  
  The uncertainty due to stochasticity in each bin is the
  standard error of the $\Delta v$ values in the bin, $\sigma_{\rm stochastic}$,
  which is calculated by

  \begin{equation} \label{over_n}
    \sigma_{\rm stochastic} ~=~ \frac{\sigma_{\Delta v(\Delta r)}}{\sqrt{n_{\rm pairs}}}
  \end{equation}

  where $\sigma_{\Delta v(\Delta r)}$ is the standard deviation of the 
  $\Delta v$ values in the bin, and $n_{\rm pairs}$ is the number of pairs
  of stars in the bin.

  If there are observational errors, then the stochastic error must use the 
  weighted standard deviation of the $\Delta v$:
  \begin{equation}
    \sigma_{\Delta v(\Delta r)} ~=~ \sqrt{\frac{\sum  w_{i\!j} (\Delta v_{i\!j} - \Delta v(\Delta r))^2}{\sum w_{i\!j}}}
  \end{equation}
  where the sums are over pairs $i\!j$ in the bin.
  
  \item \textbf{Combine the errors.} 
  
  Combine the stochastic errors with
  the observational errors calculated in step \ref{obs_err_step}
  to get the total error on $\Delta v(\Delta r)$ in each bin:
  \begin{equation}
      \sigma_{\rm total} ~=~ \sqrt{\sigma_{\rm obs}^2 ~+~ \sigma_{\rm stochastic}^2 }
  \end{equation}

  \item \textbf{Plot $\Delta v$($\Delta r$) with errorbars.} 
  
  Produce a plot
  using the magnitude definition $\Delta v_{i\!j \rm{M}}$ and  the directional
  definition $\Delta v_{i\!j \rm{D}}$. 

\end{enumerate}

As we will show, these plots contain a significant amount of quantitative and qualitative 
information on the spacial-velocity structure of a distribution.

\begin{figure}
  \includegraphics[width=\columnwidth]{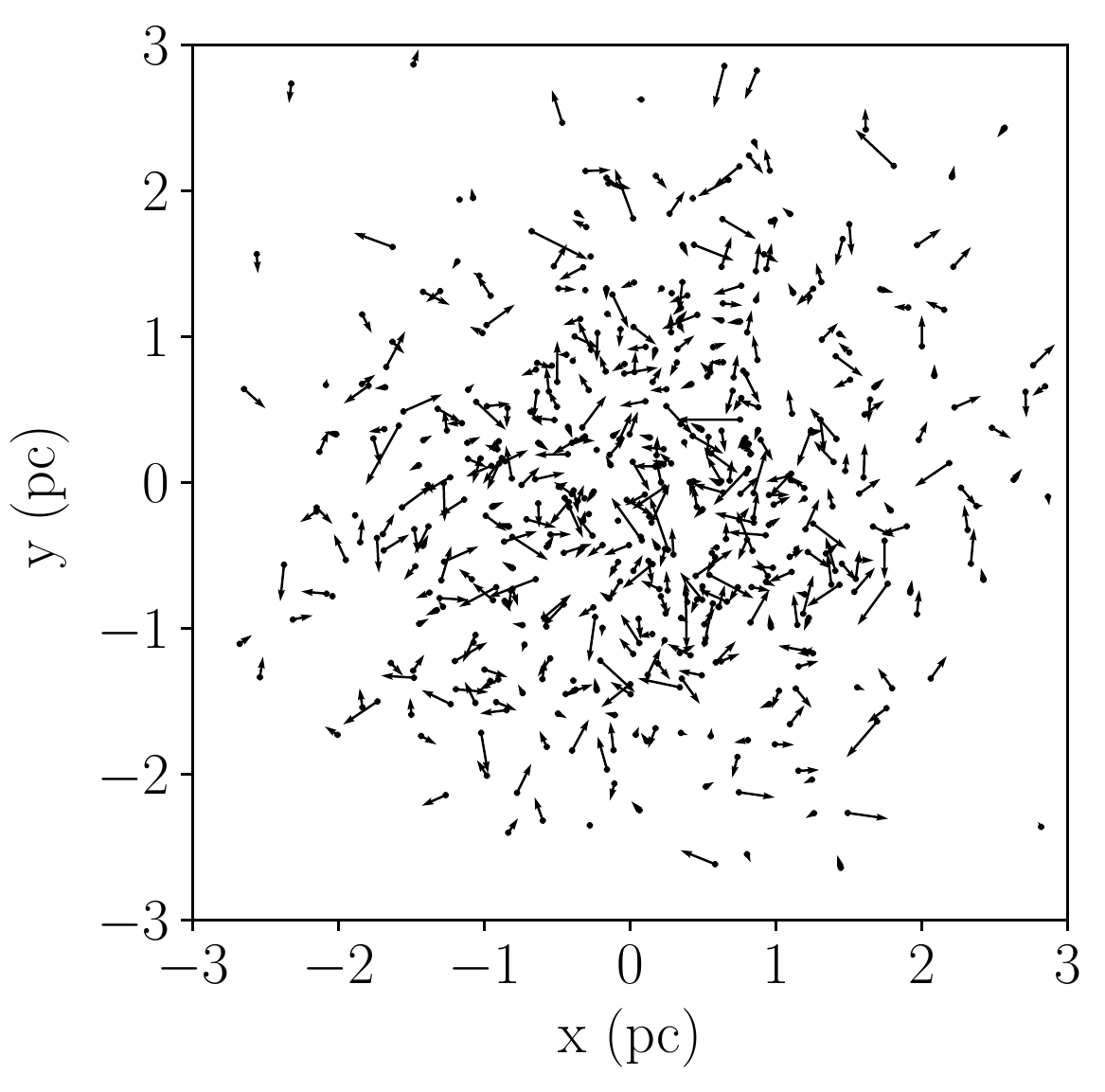}
  \caption{An artificial region with random velocities projected
    onto a 6 pc by 6 pc box in the x-y plane. Each star is represented
    by a dot with an arrow showing its velocity.}
  \label{fig_1}
\end{figure}

\begin{figure}
  \includegraphics[width=\columnwidth]{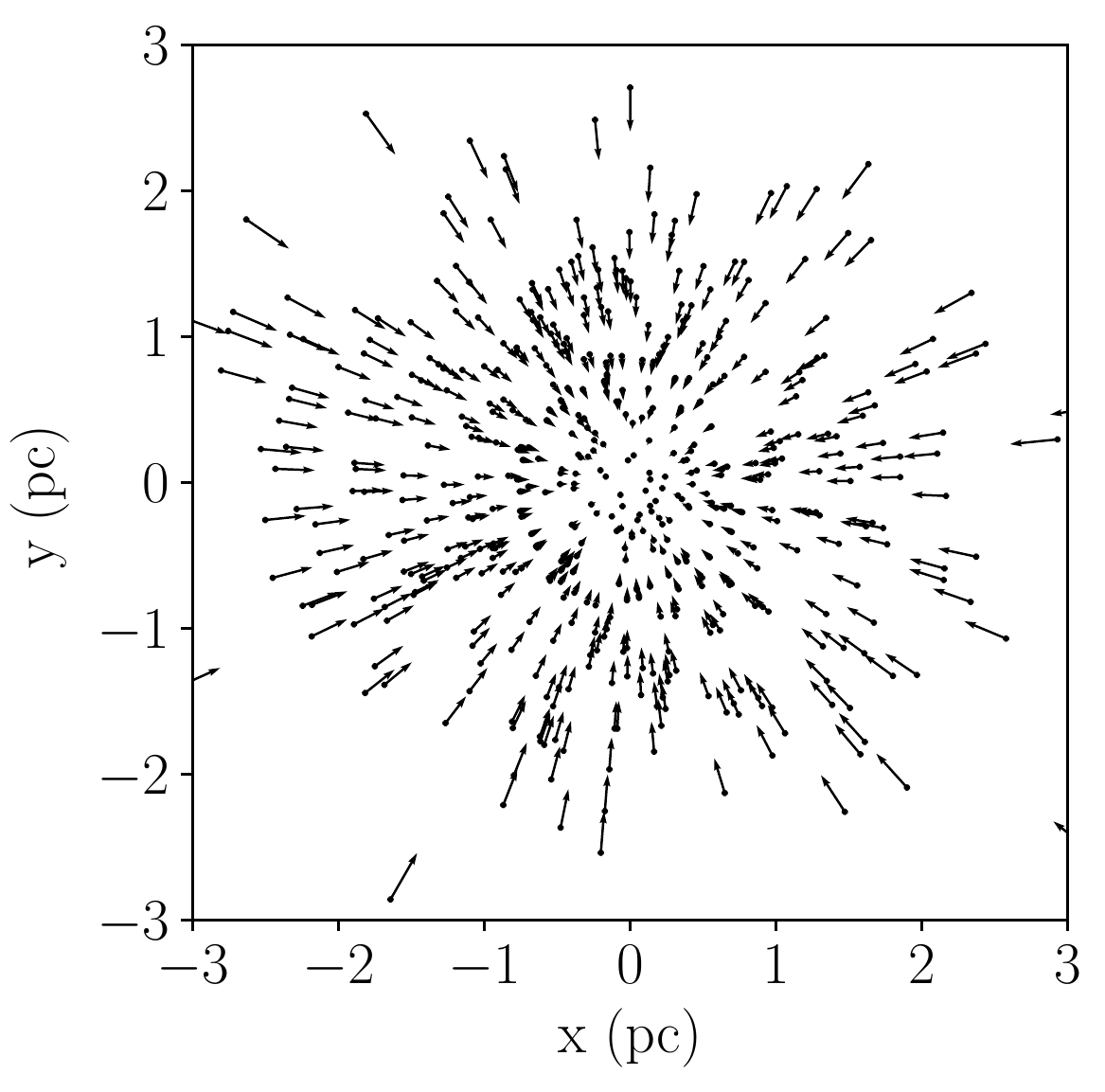}
  \caption{An artificial region projected
    onto a 6 pc by 6 pc box in the x-y plane. Each star is represented
    by a dot with an arrow showing its velocity.
   The velocity of each star is the negative of its position in order to 
   produce a very simple collapsing velocity structure.}
  \label{fig_2}
\end{figure}

To help illustrate the step by step explanation
we apply the method to two simple cases shown in Fig. \ref{fig_1} 
and Fig. \ref{fig_2}, both of which have 500 stars with guassian 
random positions. In Fig. \ref{fig_1} the velocities are also drawn 
randomly from a gaussian, so there is no correlation between a star's position 
and its velocity. In Fig. \ref{fig_2} the star's velocities are the 
negative of their positions to create a `collapsing' distribution. We provide more 
realistic examples later, but these suffice to illustrate the method.

\begin{figure}
  \includegraphics[width=\columnwidth]{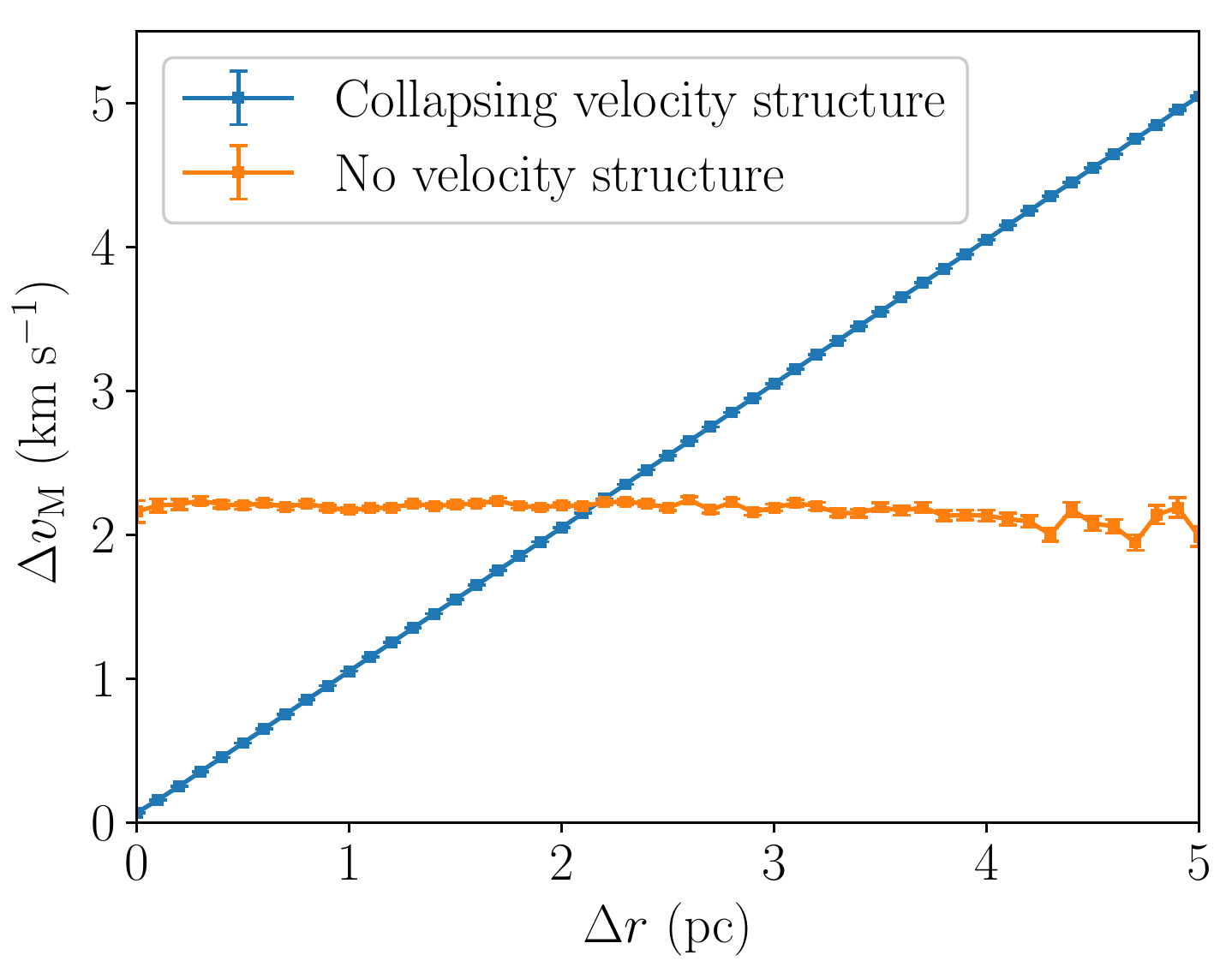}
  \caption{Physical separation, $\Delta r$, plotted against
    velocity difference as calculated by the magnitude definition,
     $\Delta v_{\rm M}$, for the region shown in Fig. \ref{fig_1}
     in orange, and for the region shown in  Fig. \ref{fig_2}
     in blue.}
  \label{fig_3}
\end{figure}

Fig. \ref{fig_3} shows $\Delta v_{\rm M}$ plotted against $\Delta r$ for 
for the random (orange line) and simple collapsing (blue line) distributions shown in Fig. 
\ref{fig_1} and Fig. \ref{fig_2}.

\begin{figure}
  \includegraphics[width=\columnwidth]{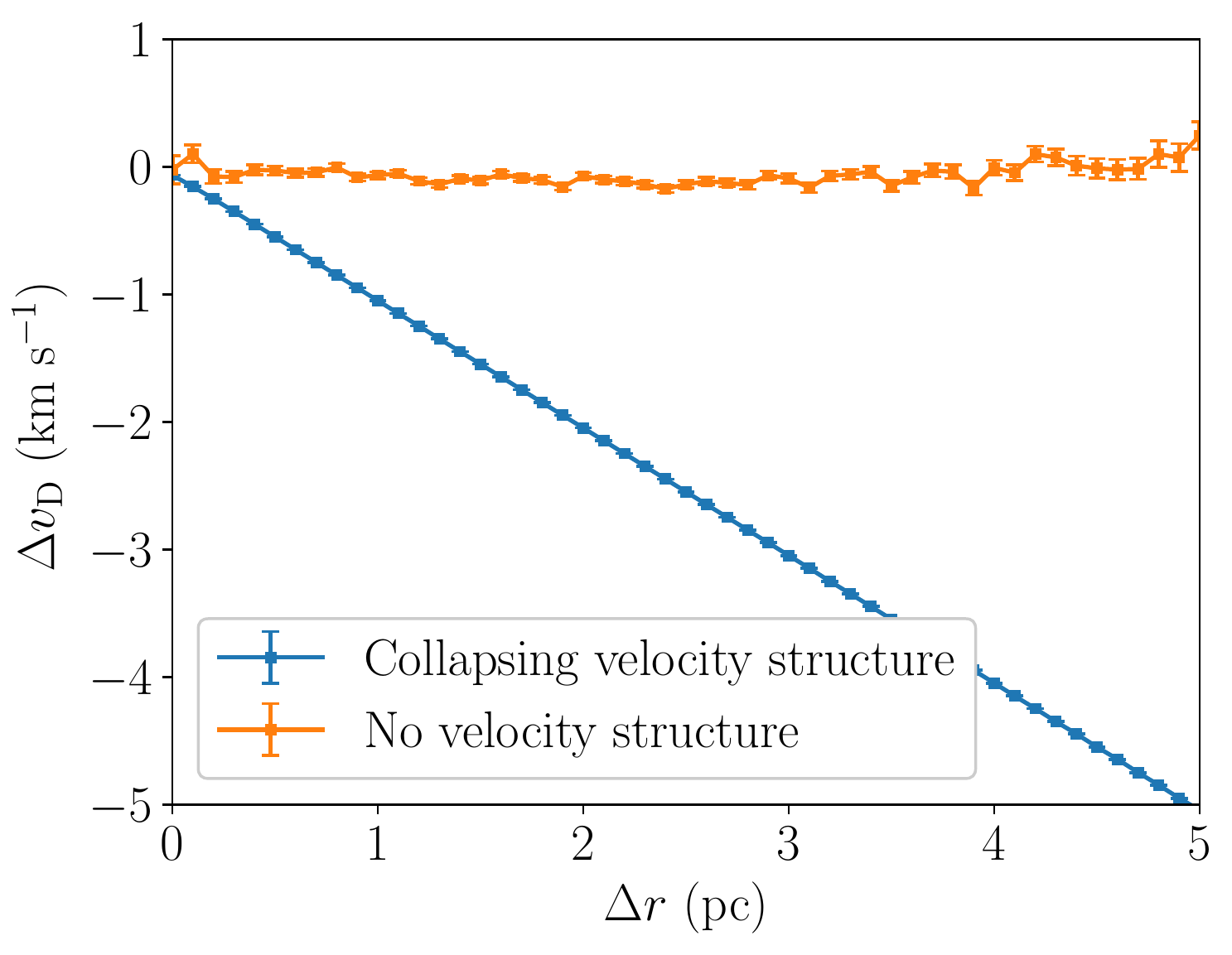}
  \caption{This plot shows physical separation, $\Delta r$, against
    velocity difference as calculated by the directional definition, 
    $\Delta v_{\rm D}$, for the region shown in Fig. \ref{fig_1}
    in orange, and for the region shown in  Fig. \ref{fig_2}
    in blue.}
  \label{fig_4}
\end{figure}

The orange line is flat which shows that in the region with random velocities 
there is no velocity structure on any spatial scale. This is as expected 
as in this region there is no correlation between the distance between two stars 
and their velocity difference. It is worth noting that from Fig. \ref{fig_1} 
the eye can be fooled into thinking that the locations of high velocity stars are 
biased towards the centre. This is an artifact of there being more stars near the 
centre, and so there is a greater chance of a high-velocity star appearing there. 
This highlights the need for objective numerical methods of analysing velocity structure.

The blue line (collapsing region) is more interesting. Because the velocities in this region
are the negative of the star's position the difference in two star's velocities is directly 
proportional to how far apart they are. Therefore we expect a linear relationship between
$\Delta r$ and $\Delta v_{\rm M}$, and this 
is clearly visible in Fig. \ref{fig_3}.
Inspection of Fig. \ref{fig_2} confirms that in this region stars
that are very close to one another (low $\Delta r$) have practically identical
velocities, so low velocity differences $\Delta v_{\rm M}$. As a result in 
Fig. \ref{fig_3} $\Delta v_{\rm M}$ is low at low $\Delta r$.
In contrast inspection of Fig. \ref{fig_2} shows that stars
that are far apart (high $\Delta r$) have very different velocities
(high $\Delta v_{\rm M}$), which is reflected in Fig. \ref{fig_3}.
      
In Fig. \ref{fig_4} $\Delta v_{\rm D}(\Delta r)$ is plotted
for the random and collapsing distributions shown in Fig. 
\ref{fig_1} and Fig. \ref{fig_2}. Recall that by this definition 
negative $\Delta v_{\rm D}$ means the stars are
moving towards one another, and positive $\Delta v_{\rm D}$ means the stars are
moving apart.

For the random velocity distribution (orange line) $\Delta v_{\rm D}(\Delta r)$ is flat, 
again showing no preferred scales or trends.  It has a value of roughly zero showing
no global expansion or contraction as expected given the velocities were drawn from a gaussian
distribution centred on zero.
  
The blue line (collapsing distribution) is entirely negative indicating
that at all separations stars are moving towards each other. Again, given
that this region is collapsing that is expected. We also see that $\Delta v_{\rm D}$
becomes more negative as $\Delta r$ increases. This is because stars
have that are further apart are moving towards each other faster in this region.

We draw the readers attention to the increase
in the error with $\Delta r$ visible
in Fig. \ref{fig_3} and Fig. \ref{fig_4}.
This is due to the decreasing number of pairs in bins with larger and larger 
$\Delta r$. As a result $n_{\rm pairs}$ is low for very
high $\Delta r$ bins and from Eqn. \ref{over_n} the uncertainties are larger.

\section{Plummer spheres} \label{Applying the method to a Plummer sphere}

The examples used above are very simplistic. In 
this section we apply the method to the more
realistic case of Plummer spheres. 

We generate a Plummer sphere using the method of \citet*{Aarseth74}, 
with 1000 stars and a half mass radius of 2 pc. 
We scale the velocities by three different factors to produce one Plummer sphere with 
virial ratio $\alpha_{\rm vir}$ = 0.3 (sub-virial), one with
$\alpha_{\rm vir}$ = 0.5 (virialised), and one with $\alpha_{\rm vir}$ = 0.7 (super-virial).
Here $\alpha_{\rm vir} = T/|\Omega|$, where $T$ is the kinetic energy, 
and $\Omega$ is the potential energy.

We would expect a sub-virial distribution to collapse and a super-virial distribution to 
expand but we have not imposed this in any way other than by scaling {\em all} velocities 
by the appropriate factor.  We run an $N$-body simulation of each Plummer sphere for 1~Myr 
in order to allow them to start to adapt to the imposed virial ratios.

\begin{figure}
  \includegraphics[width=\columnwidth]{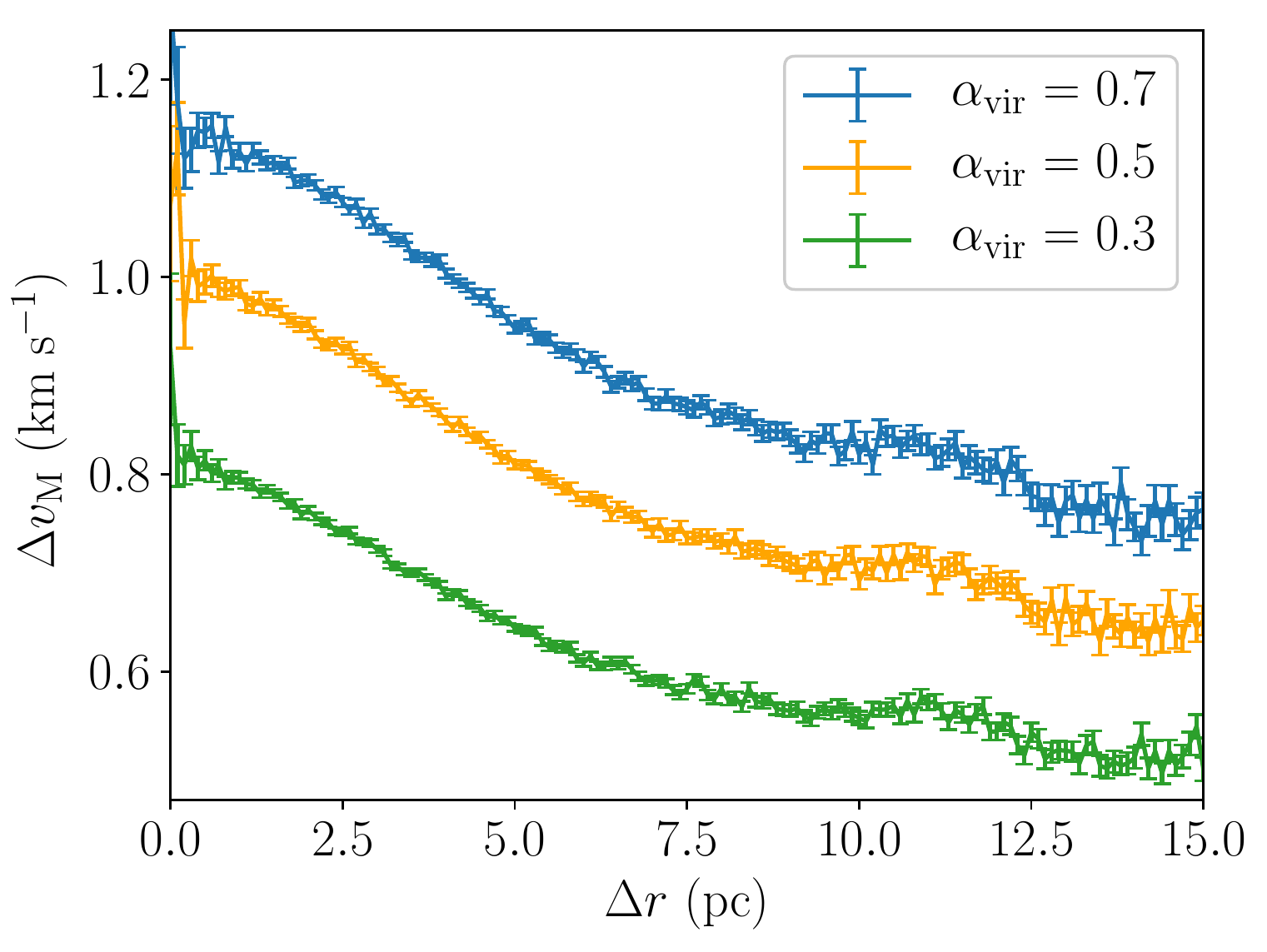}
  \caption{Plot showing $\Delta v_{\rm M}(\Delta r)$ for three Plummer spheres. The x axis is
    physical separation $\Delta r$ and the y axis is velocity difference $\Delta v_{\rm M}$.
    $\Delta v_{\rm M}(\Delta r)$ of the $\alpha_{\rm vir}$ = 0.7 case is shown by a blue line,
    the $\alpha_{\rm vir}$ = 0.5 case by an orange line, and the $\alpha_{\rm vir}$ = 0.3 case by a
    green line.}
  \label{fig_5}
\end{figure}

\begin{figure}
  \includegraphics[width=\columnwidth]{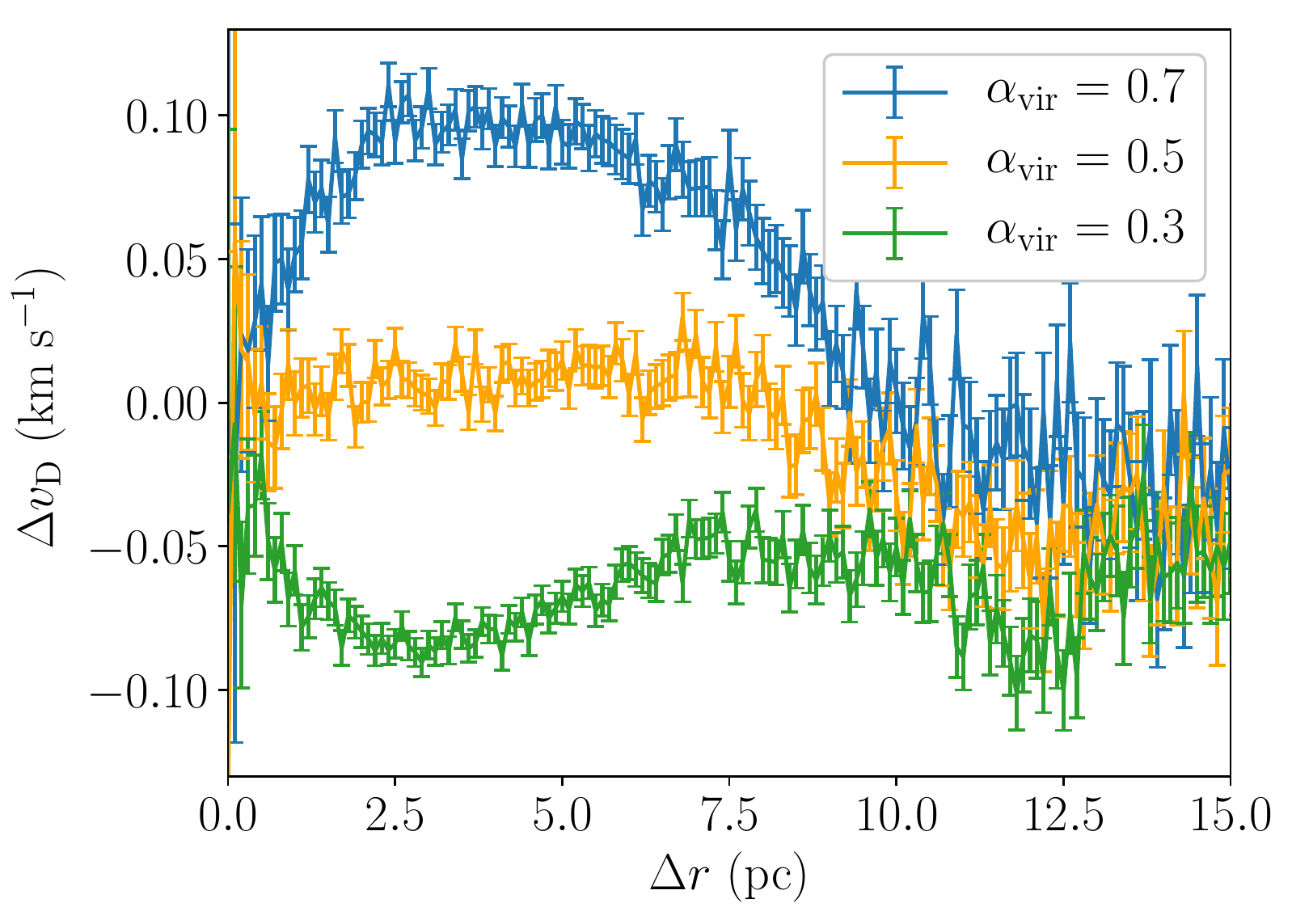}
  \caption{Plot showing $\Delta v_{\rm D}(\Delta r)$ for three Plummer spheres. The x axis is
  physical separation $\Delta r$ and the y axis is velocity difference $\Delta v_{\rm D}$.
  $\Delta v_{\rm D}(\Delta r)$ of the $\alpha_{\rm vir}$ = 0.7 case is shown by a blue line,
  the $\alpha_{\rm vir}$ = 0.5 case by an orange line, and the $\alpha_{\rm vir}$ = 0.3 case by a
  green line.}
  \label{fig_6}
\end{figure}

Fig. \ref{fig_5} shows $\Delta v_{\rm M}(\Delta r)$, and 
Fig. \ref{fig_6} $\Delta v_{\rm D}(\Delta r)$ for all three 
Plummer spheres. In both figures the green lines are used for the $\alpha_{\rm vir} =0.3$ 
Plummer sphere, orange for the $\alpha_{\rm vir} =0.5$ Plummer sphere, and blue for the $\alpha_{\rm vir} =0.7$ Plummer sphere.

In Fig. \ref{fig_5} all three lines have the same shape: 
large $\Delta v_{\rm M}$ at low $\Delta r$ which decreases towards high $\Delta r$.  
The reason for this is that Plummer spheres have a high central velocity dispersion 
(at the deepest part of the potential) which decreases at larger radii.
The majority of pairs of stars with low $\Delta r$ are located
in the core as, by definition, this area is dense and so contains
many stars that are close together. These low $\Delta r$ pairs are 
therefore made up of stars with a high velocity dispersion so any two star's velocity vectors
are likely to be very different, and the magnitude of this difference, $\Delta v_{\rm M}$, will be large.  
In contrast stars that make up high $\Delta r$ pairs are predominantly located in the halo 
where the velocity dispersion is smaller so $\Delta v_{\rm M}$ is low.

There is a clear vertical offset between Plummer spheres with higher virial ratios in this figure. This is because, as virial
ratio is the ratio of kinetic to potential energies, stars in regions
with high virial ratios will have higher speeds on average.
Therefore velocity differences
between pairs of stars in those regions are more likely to be high.

Otherwise the velocity structures of the three Plummer spheres
are near identical according to $\Delta v_{\rm M}$. There is a `kink' present 
in all three lines at $\Delta r$ $\sim 11$ pc. This is just a peculiar 
feature of this particular Plummer sphere realisation (a similar feature is not 
present in Plummer spheres generated with different random number seeds).

In Fig. \ref{fig_6} we show $\Delta v(\Delta r)$ for each of the Plummer spheres
using the directional definition $\Delta v_{\rm D}$.  
While in Fig. \ref{fig_5} all three Plummer spheres showed 
the same velocity structure with only a vertical offset due to their virial ratio, here 
the three Plummer spheres appear quite different.

Recall that positive $\Delta v_{\rm D}$ is indicative of expansion, and a negative value 
is indicative of collapse.  The blue line ($\alpha_{\rm vir} =0.7$) has values 
that are generally positive, the orange line ($\alpha_{\rm vir} =0.5$) is roughly 
flat, and the green line ($\alpha_{\rm vir} =0.3$) is always negative.

We examine the $\alpha_{\rm vir} =0.5$ Plummer sphere (orange line) first.  
For separations of less than 10 pc (i.e. separations that contain the majority of the pairs of stars), 
$\Delta v_{\rm D}$ is flat, showing that stars are equally likely to be moving towards 
each other as away from each other. This is as would be expected for a region that is in neither bulk expansion nor contraction.
At separations above 10 pc the stars are generally moving towards
each other. This may be due to stars with such extreme
separations being mainly found in the extreme halo of the Plummer sphere, and
they are being attracted back towards the centre. As a result
they are moving towards each other on average. However given the
size of the error bars it is also possible the apparent inconsistency of the
velocity structure with zero at large separations is an artefact of stochasticity.

For the collapsing ($\alpha_{\rm vir}$ = 0.3) case the line is below zero
at every separation. That means at every separation on average the
stars are moving closer together.

The expanding case has positive $\Delta v_{\rm D}$ at separations
below $\sim$10 pc; on average stars at these separation
are moving away from each other. As in the $\alpha_{\rm vir}$ = 0.5 case
stars with extreme separations are found to be moving towards one
another. Again this may be due to stars on the outskirts being attracted back towards
the centre or it may due to a combination of stochasticity
and large error bars at high $\Delta r$. 

\begin{figure}
  \includegraphics[width=\columnwidth]{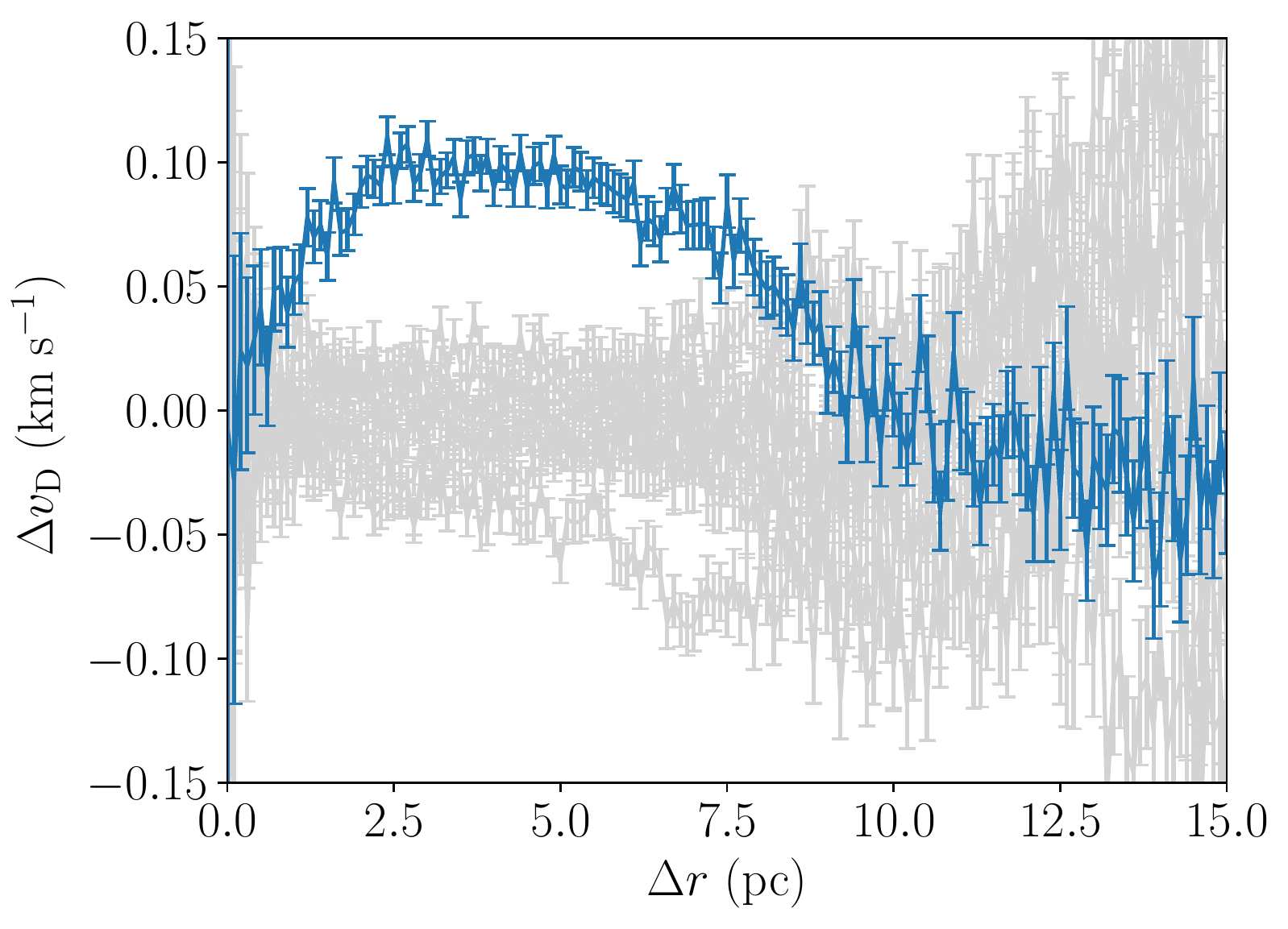}
  \caption{Plot showing $\Delta v_{\rm D}(\Delta r)$ for the $\alpha_{\rm vir} =0.7$ Plummer sphere. 
  he true velocity structure is shown in blue. After the velocities are randomly shuffled between stars 
  the recalulated velocity structure is plotted in grey.
  The x axis is
  physical separation $\Delta r$ and the y axis is velocity difference $\Delta v_{\rm D}$.}
  \label{fig_7}
\end{figure}

Uncertainty over whether a feature is real or an `artefact' can be an issue
in bins where $n_{\rm pairs}$ is low, as is typically the case in large $\Delta r$ bins.
To examine whether this feature is significant velocities are shuffled randomly 
between stars which removes any real velocity structure from the data. The method 
is then reapplied and any `features' observed in the result must be due to stochasticity. 
This is done ten times and the results are plotted in grey in 
Fig. \ref{fig_7}. The actual velocity structure is again plotted in blue for comparison.

From Fig. \ref{fig_7} it is is clear that any `features' in the actual 
velocity stucture of the $\alpha_{\rm vir} =0.7$ Plummer sphere at $\Delta r >$ 
$\sim 9$ pc are not significant. The same analysis is applied to the $\alpha_{\rm vir} =0.3$ 
and $\alpha_{\rm vir} =0.5$ Plummer spheres. In the $\alpha_{\rm vir} =0.3$ case all
features are found to be significant up to $\Delta r \sim 13$ pc, and in the $\alpha_{\rm vir} =0.5$
case the structure is found to be consistent with the randomised cases (so no systematic expansion or
contraction) at all $\Delta r$s.

\subsection{Interpreting observations}

If an observer observed the three spherical clusters in this section they would find
them to be very similar in their spacial structure.
An analysis of their velocity magnitudes 
$\Delta v_{\rm M}$ would show a structure such as in Fig. \ref{fig_5} 
and it would be possible to say that they each have a Plummer-like 
velocity distribution. Additionally, an analysis of 
$\Delta v_{\rm D}(\Delta r)$ would show that one is expanding, another collapsing, 
and the other appears static.

\section{Complex substructured regions} \label{fractals}

Plummer spheres are fairly simple example distributions.  We now 
apply the method to complex substructured distributions generated by the box fractal method.

A full description of the box fractal method is available
in \citet{Goodwin04}, however a brief overview is given here.
A single `parent'
star is placed in the centre of a box, and then the box
is divided into smaller boxes. 
The probability that each of these smaller boxes
has of containing a `child' star is chosen by the user. If the probability is low the  
fractal will have a high degree of substructure, and if the probability is large the fractal will be more smooth.
If a box does contain a child star it is placed approximately
in the centre of the box (noise is added
to the position to avoid an obviously gridlike structure).
The velocity of the child star is the same as its parent's
velocity plus some random component. 
After this each child star becomes a parent and the process is
repeated to produce the desired number of stars (extra stars can be deleted at random).

Note that here we are only interested in investigating the application of the VSAT method
to substructured distributions, 
so the absolute values of e.g. radius and virial ratio are unimportant.

\subsection{Distributions with low substructure} \label{Low substructured regions}

\begin{figure}
  \includegraphics[width=\columnwidth]{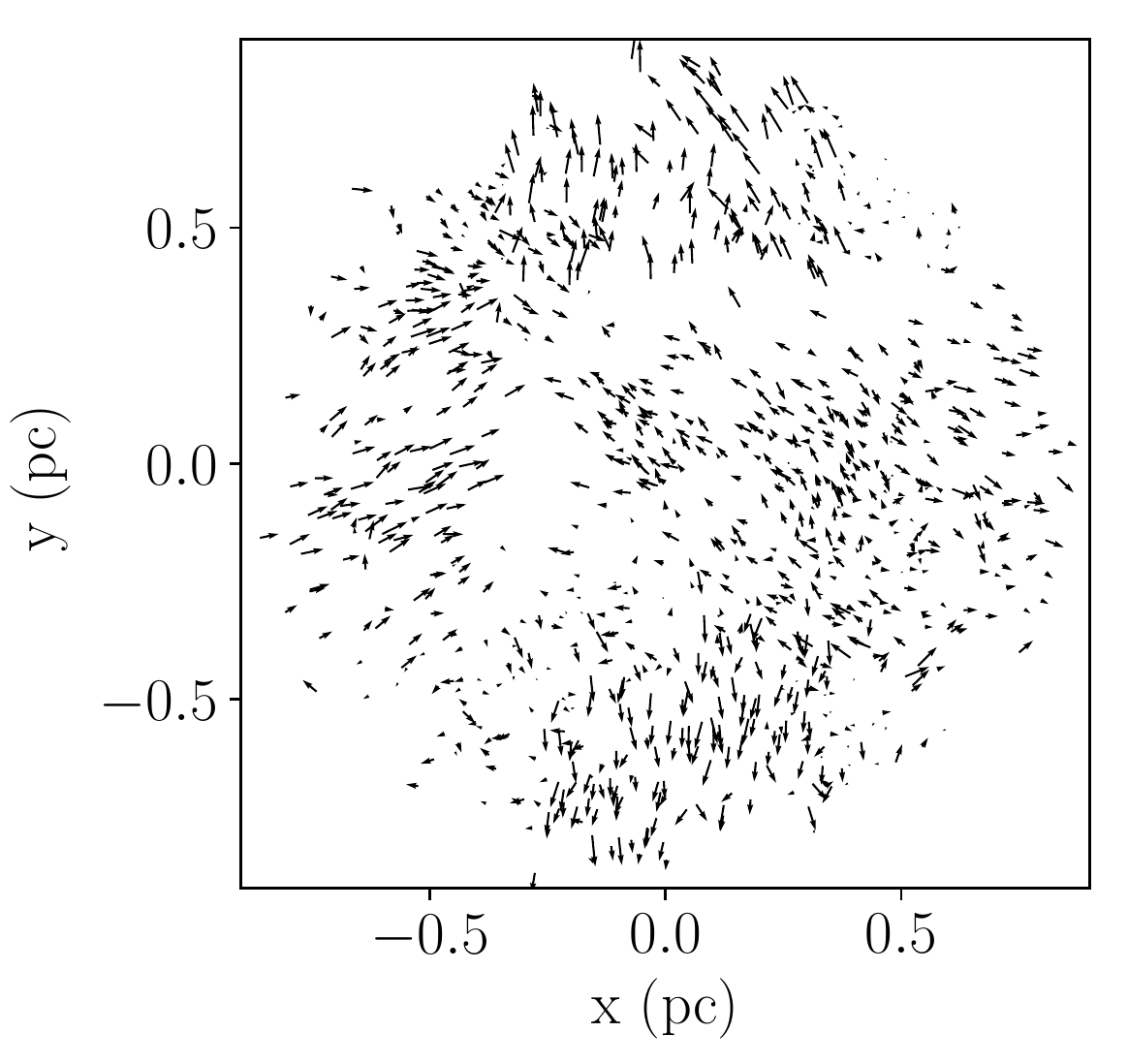}
  \caption{A distribution with low substructure
    generated by the box fractal method
    projected into a 1.8 pc $\times$ 1.8 pc box. Each star is represented
    by an arrow. The position of the arrow corresponds
    to the position of the star and the arrow itself indicates
    the star's velocity.}
  \label{fig_8}
\end{figure}

An example of a fractal with low substructure and 1000 stars is shown 
in Fig. \ref{fig_8} and the arrows show 2D velocity vectors. 
Clear structure in both the positions and velocities of the stars is 
obvious, but too complex to interpret by eye in any meaningful way. It
is possible to tell there is substructure, but without other information the eye could not reasonably judge the degree of velocity substructure or
how it is distributed.

\begin{figure}
  \includegraphics[width=\columnwidth]{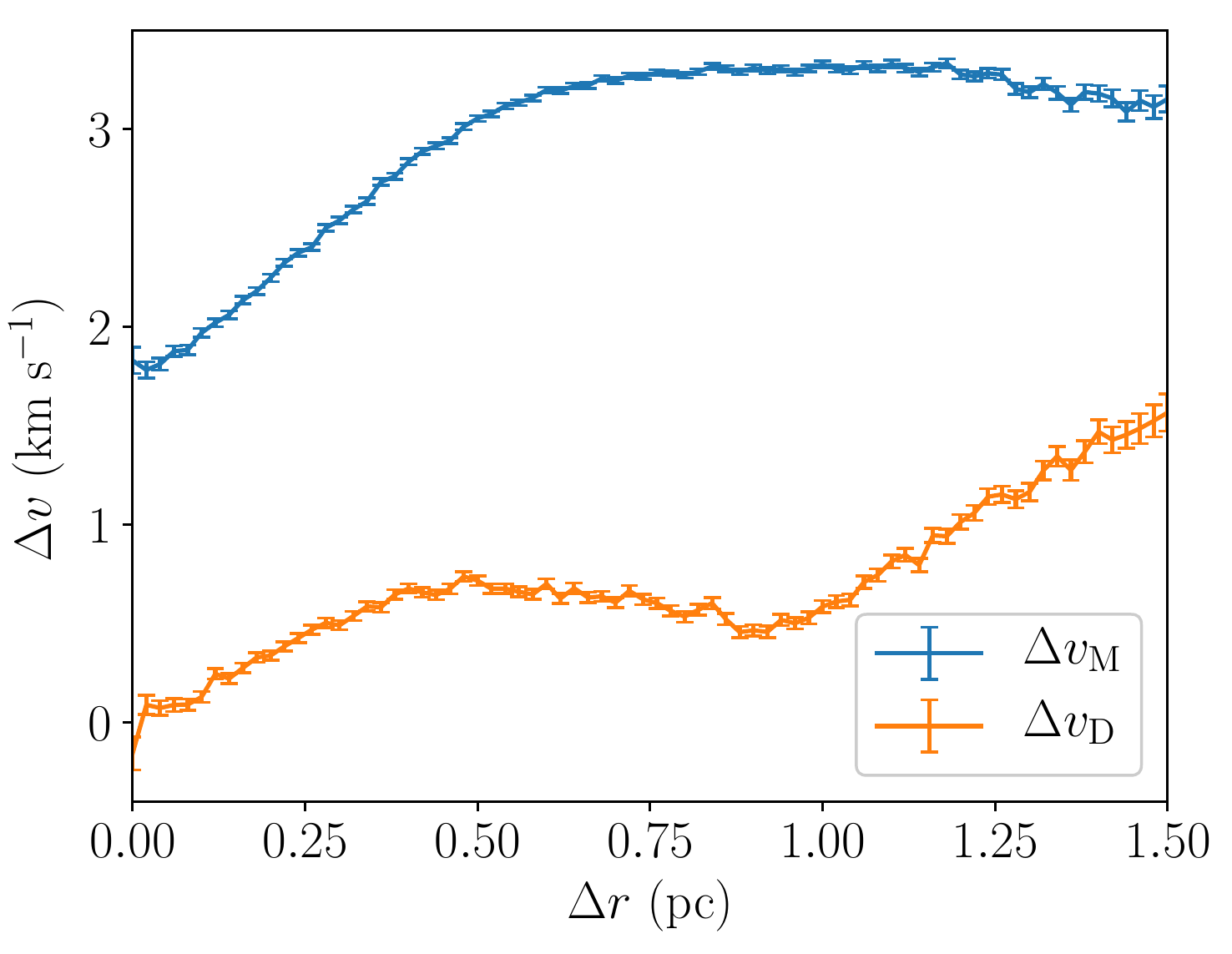}
  \caption{The velocity structure of the distribution with low substructure shown 
  in Fig. \ref{fig_8}. The velocity structure $\Delta v_{\rm M}(\Delta r)$
  is shown by a blue line and $\Delta v_{\rm D}(\Delta r)$ by 
  an orange line.}
  \label{fig_9}
\end{figure}

In Fig. \ref{fig_9} we show the magnitude (blue line)
and directional (orange line) $\Delta v(\Delta r)$ plots 
for the fractal in Fig. \ref{fig_8} (note that everything is done in 2D).
  
$\Delta v_{\rm M}(\Delta r)$ (blue line), is $\sim 2$ km s$^{-1}$ when the separations 
are low, rising to $\sim 3$ km s$^{-1}$ at separations of $\sim 0.7$ pc and then remaining roughly constant. 

This initial increase of $\Delta v_{\rm M}$ with $\Delta r$ is because, as
described above, when child stars are produced they inherit most
of their velocities from their parents, plus a random
component. As a result
in the completed distributions the stars closest together have
very similar velocity vectors, so the magnitude of their difference, $\Delta v_{\rm M}$, 
is small. Stars further away from each other are very distantly `related' so 
have very different velocity vectors and their $\Delta v_{\rm M}$ is big.

The 0.7 pc length scale is significant because it is the approximate
radius of the distribution. Stars separated by this length scale or greater
are generated from different `child stars' of the very first generation
in the production of the fractal. The random changes applied at each generation after that
average to a net additional difference of zero, so $\Delta v_{\rm M}$ remains roughly flat
at $\Delta r \geq$  0.7 pc.

The directional velocity structure, $\Delta v_{\rm D}(\Delta r)$ (orange line), 
is always positive meaning that stars tend to move away from each other on all scales.  
There is some structure in $\Delta v_{\rm D}$ showing that expansion increases 
on scales up to 0.5 pc, then is roughly even, before increasing again on scales of $> 1$ pc.

\begin{figure}
  \begin{subfigure}[b]{\columnwidth}
          \centering
          \includegraphics[width=\columnwidth]{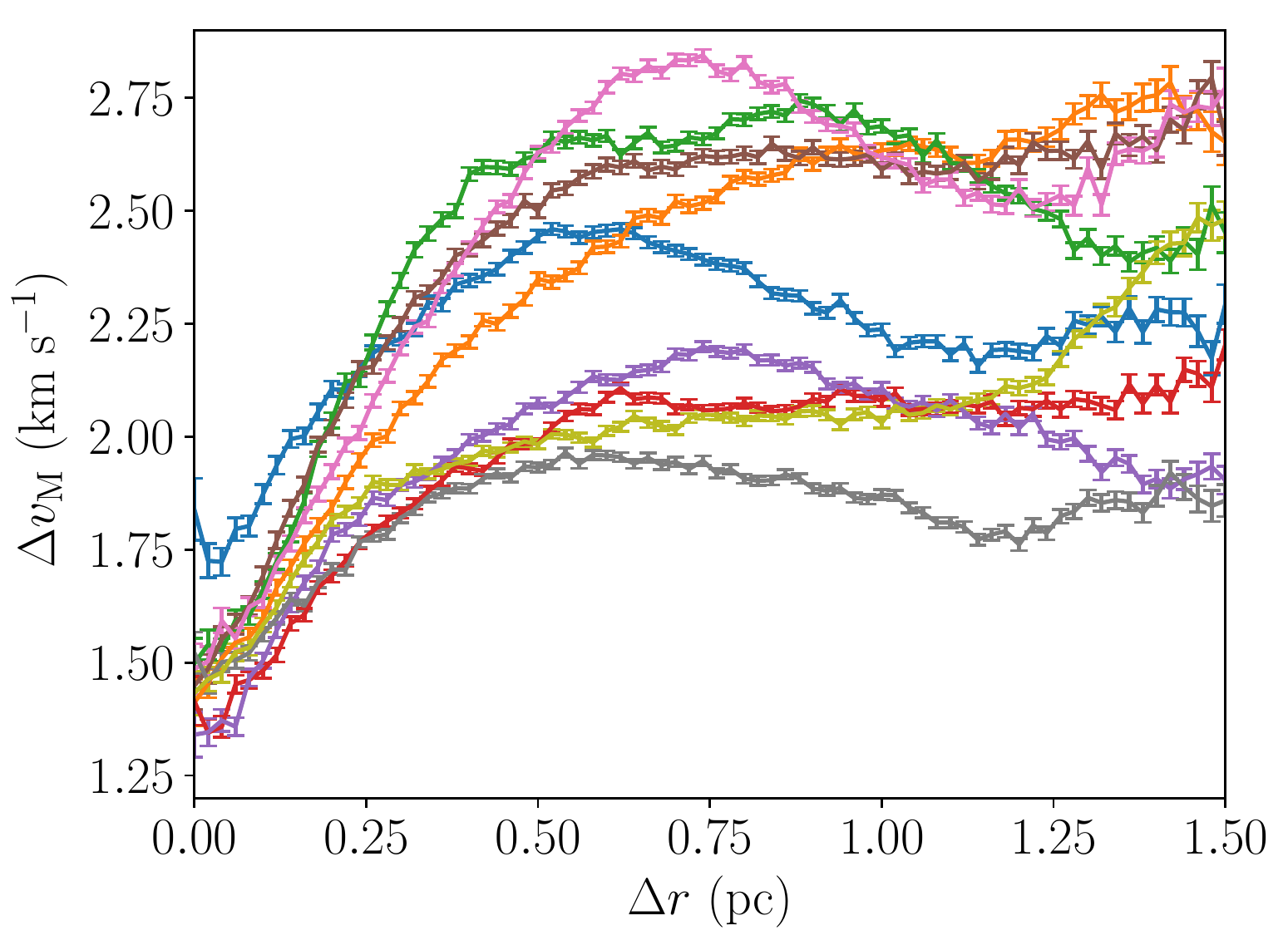}
          \label{fig_10_a}
  \end{subfigure}
  
  \begin{subfigure}[b]{\columnwidth}
          \centering
          \includegraphics[width=\columnwidth]{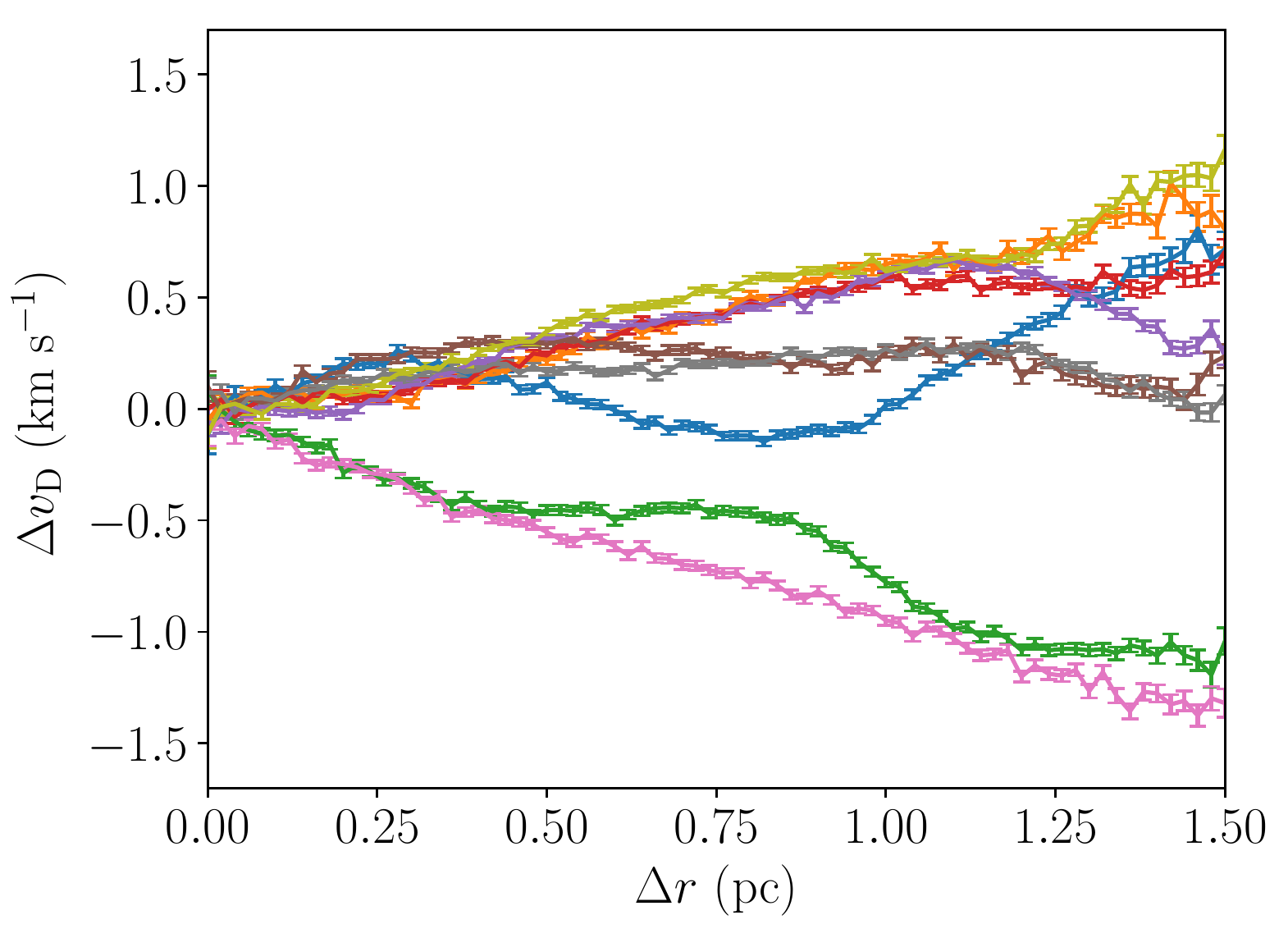}
          \label{fig_10_b}
  \end{subfigure}
  \caption{This figure shows $\Delta v_{\rm M}(\Delta r)$ (top panel) 
  and $\Delta v_{\rm D}(\Delta r)$ (bottom panel) of nine
  distributions with low substructure generated by the box fractal method.}
  \label{fig_10}
\end{figure}

Fig. \ref{fig_10} shows $\Delta v_{\rm M}(\Delta r)$ (top panel) 
and $\Delta v_{\rm D}(\Delta r)$ (bottom panel) for nine distributions statistically 
identical to that in Fig. \ref{fig_8} (only the random number 
seed used to generate the distributions has been changed).  
Each distribution has the same colour in both panels.

In the top panel of Fig. \ref{fig_10} every distribution's 
velocity magnitude structure has the same basic shape: low 
$\Delta v_{\rm M}$ at small separations which increases with 
separation to up to around 0.7 pc and then is roughly flat.  
That said, the details of each individual line (distribution) are different, 
and some show `structure' at larger scales.  

In the bottom panel of Fig. \ref{fig_10} some distributions have
predominantly negative (collapsing) $\Delta v_{\rm D}$ and some predominantly positive (expanding) because
the box fractal method does not preferentially make either expanding or collapsing 
distributions. There are features visible on individual lines in this plot, reflecting
that individual distributions (and parts of individual distributions)
do have some velocity structure.

\subsection{Highly substructured distributions} \label{Highly substructured regions}

\begin{figure}
  \includegraphics[width=\columnwidth]{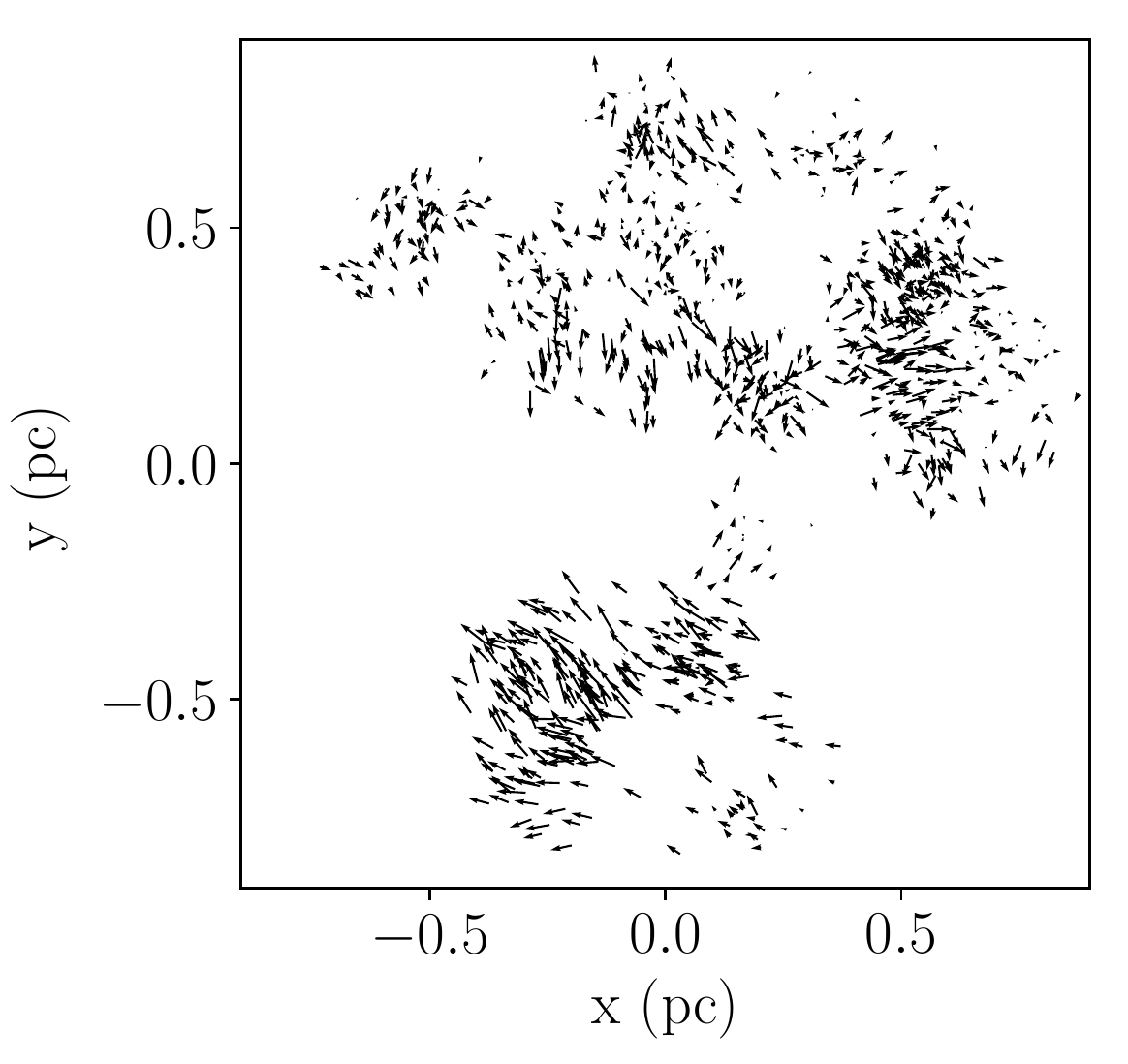}
  \caption{A highly substructured distribution generated by the box fractal method
  projected into a 1.8 pc $\times$ 1.8 pc box. Each star is represented
  by an arrow. The position of the arrow corresponds
  to the position of the star and the arrow itself indicates
  the star's velocity.}
  \label{fig_11}
\end{figure}

We now examine in detail a distribution with high substructure, illustrated in  
Fig. \ref{fig_11}, again with arrows showing the 2D velocities.  
This distribution has very clear spacial and velocity structure on a variety of scales.
Highly substructured distributions are produced using the box fractal
method by reducing the probability of each box containing a `child' star. The resulting
distribution is less smooth as stars only continue to be generated 
in boxes that do have children.

\begin{figure}
  \includegraphics[width=\columnwidth]{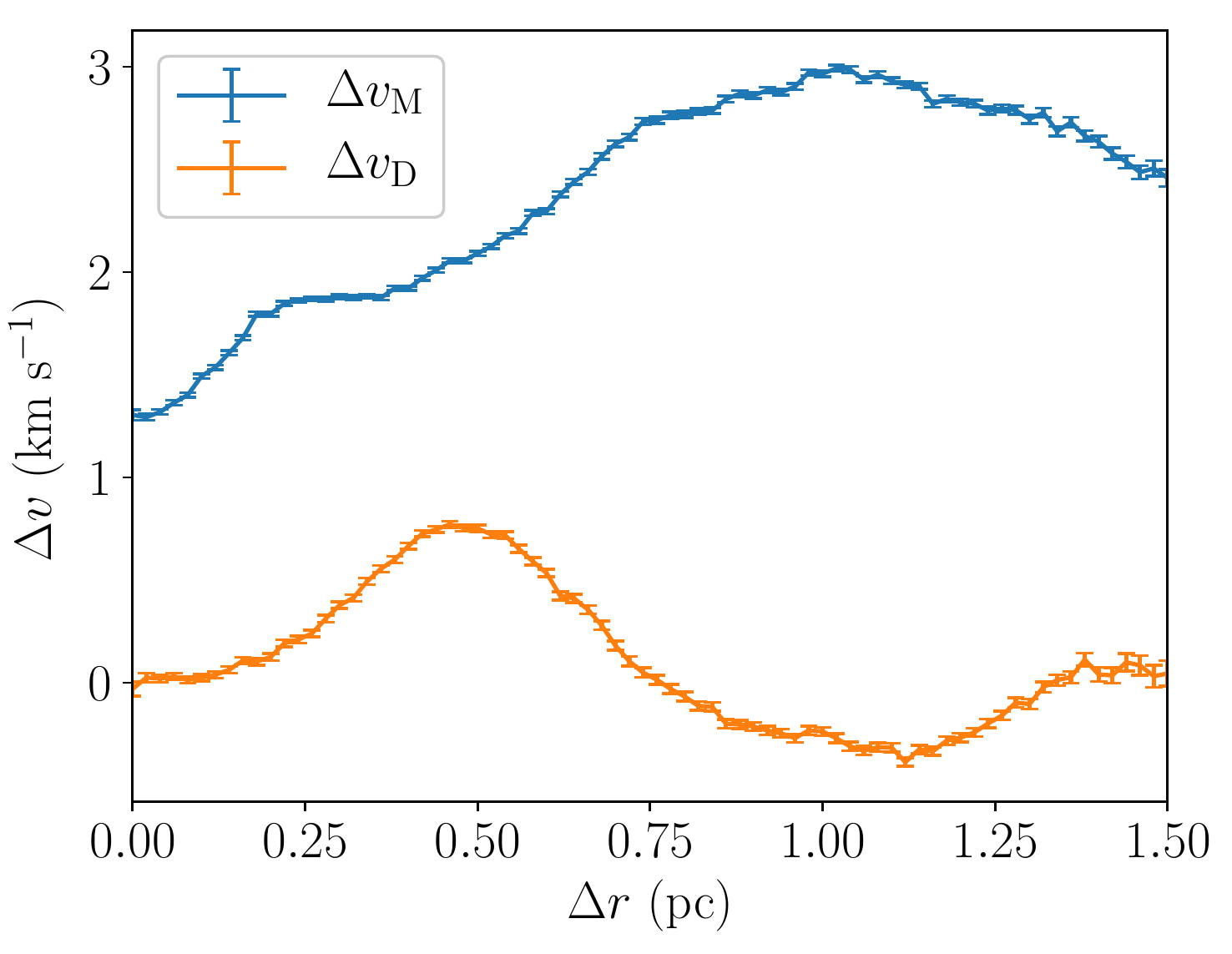}
  \caption{The velocity structure of the distribution with high substructure shown 
  in Fig. \ref{fig_11}. The velocity structure $\Delta v_{\rm M}(\Delta r)$
  is shown by a blue line and $\Delta v_{\rm D}(\Delta r)$ by 
  an orange line.}
  \label{fig_12}
\end{figure}

The velocity structure of the highly substructured distribution from Fig. \ref{fig_11} is 
shown in Fig. \ref{fig_12}, where $\Delta v_{\rm M}(\Delta r)$
is shown by the blue line and $\Delta v_{\rm D}(\Delta r)$ is shown in 
orange.

\begin{figure}
  \begin{subfigure}[b]{\columnwidth}
          \centering
          \includegraphics[width=\columnwidth]{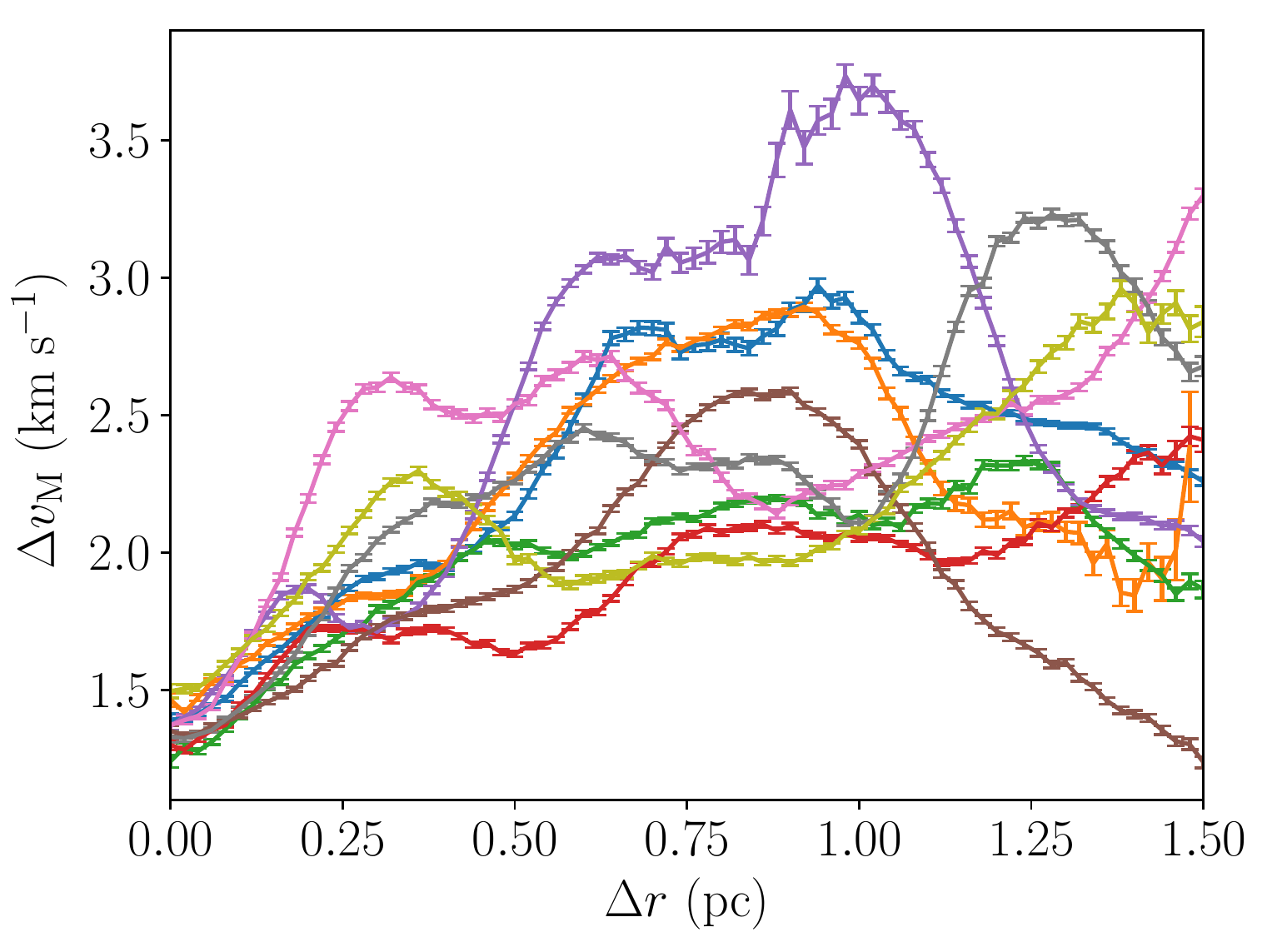}
          \label{fig_14_a}
  \end{subfigure}
  
  \begin{subfigure}[b]{\columnwidth}
          \centering
          \includegraphics[width=\columnwidth]{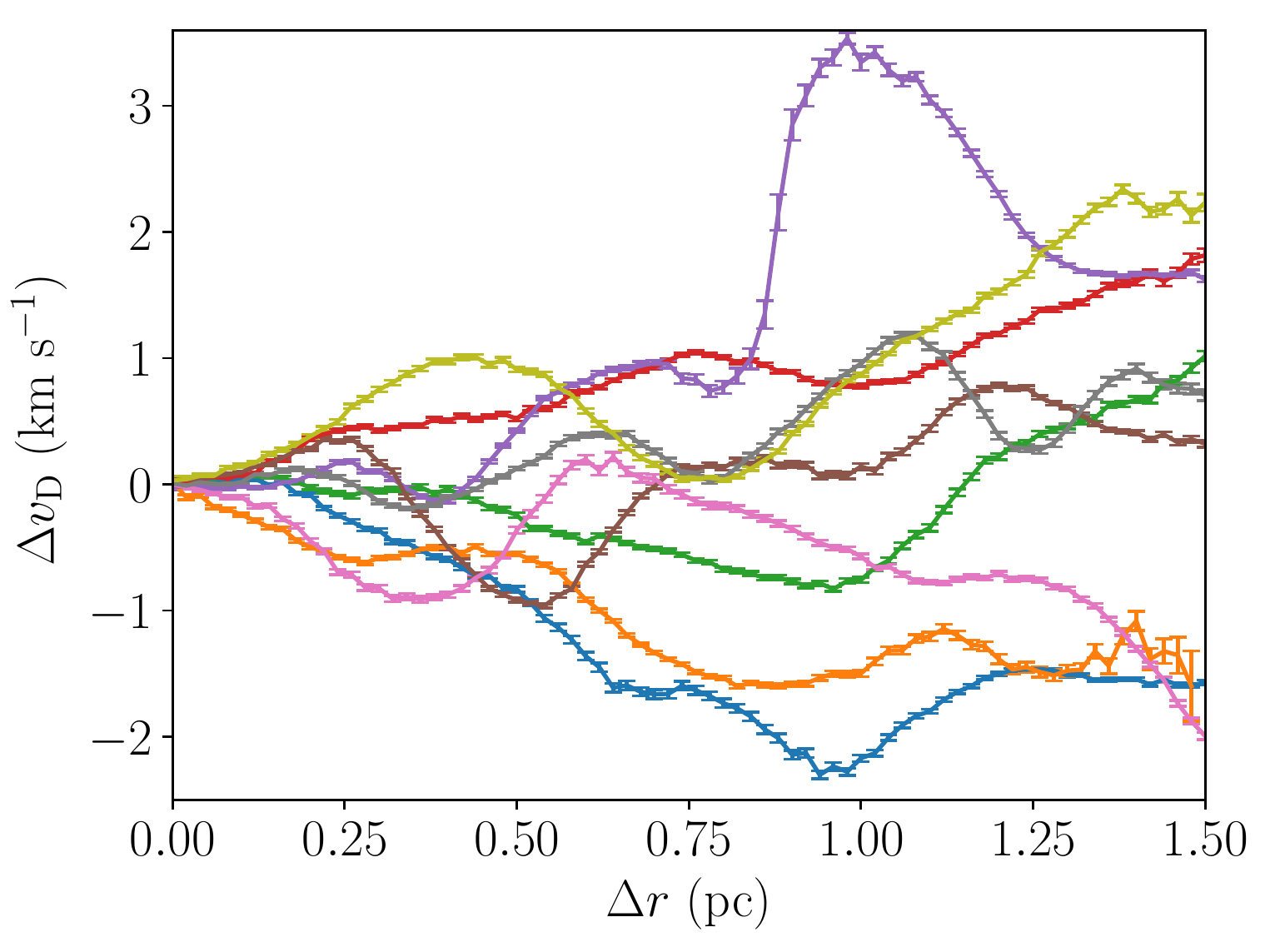}
          \label{fig_14_b}
  \end{subfigure}
  \caption{This figure shows the velocity structure $\Delta v_{\rm M}(\Delta r)$ 
  (top panel) and $\Delta v_{\rm D}(\Delta r)$
    (bottom panel) of nine
     highly substructured distributions generated by the box fractal method.}
  \label{fig_14}
\end{figure}

Broadly, the $\Delta v_{\rm M}(\Delta r)$ of the highly substructured distribution has
the same shape as $\Delta v_{\rm M}(\Delta r)$ of the distribution with low
substructure: $\Delta v_{\rm M}$ increases with $\Delta r$ and then plateaus.
However, as we would expect, in the highly substructured case the line has additional features,
including a plateau at $\sim$ 0.3 pc and a dip at $\Delta r >$ 1.1 pc. As will be shown
in Fig. \ref{fig_14} and discussed later the features in $\Delta v(\Delta r)$
due to velocity substructure are often significant in the highly substructured distributions.

The $\Delta v_{\rm D}(\Delta r)$ of the distribution
in Fig. \ref{fig_11} will now be examined in detail 
in order to demonstrate using the velocity structure plots to investigate
the detailed dynamical structure of a region (recall that this is the orange line in Fig.
\ref{fig_12}). Inspection of this figure
shows a `peak' in $\Delta v_{\rm D}$ between 
$\Delta r \sim$ 0.3 and $\Delta r \sim$ 0.6 pc, and a `trough' in $\Delta v_{\rm D}$ between 
$\Delta r \sim$ 0.8 and $\Delta r \sim$ 1.2 pc.

\begin{figure}
         \begin{subfigure}[b]{\columnwidth}
                 \centering
                 \includegraphics[width=\columnwidth]{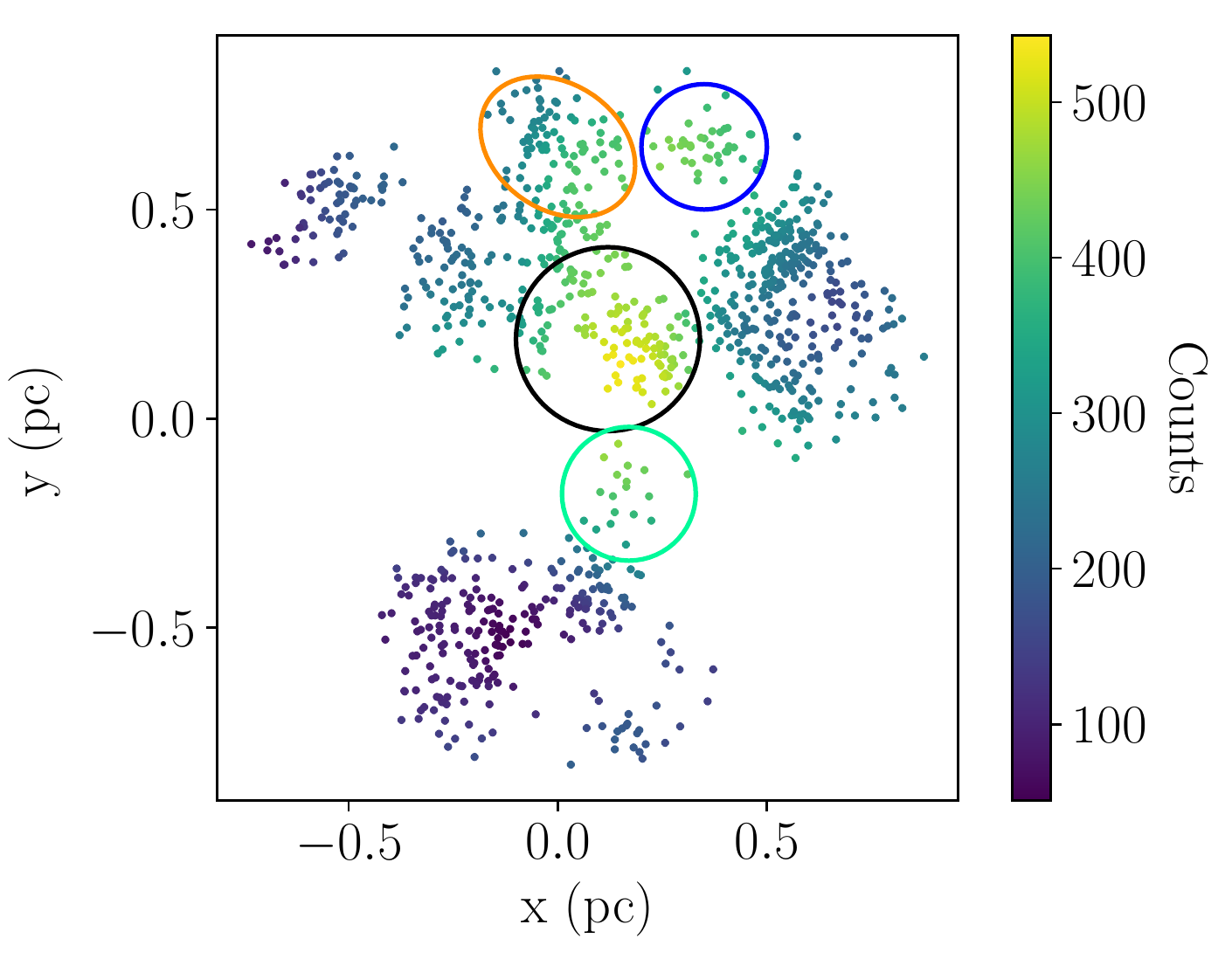}
                 {\phantomsubcaption\label{fig_13_a}}
         \end{subfigure}
         
         \begin{subfigure}[b]{\columnwidth}
                 \centering
                 \includegraphics[width=\columnwidth]{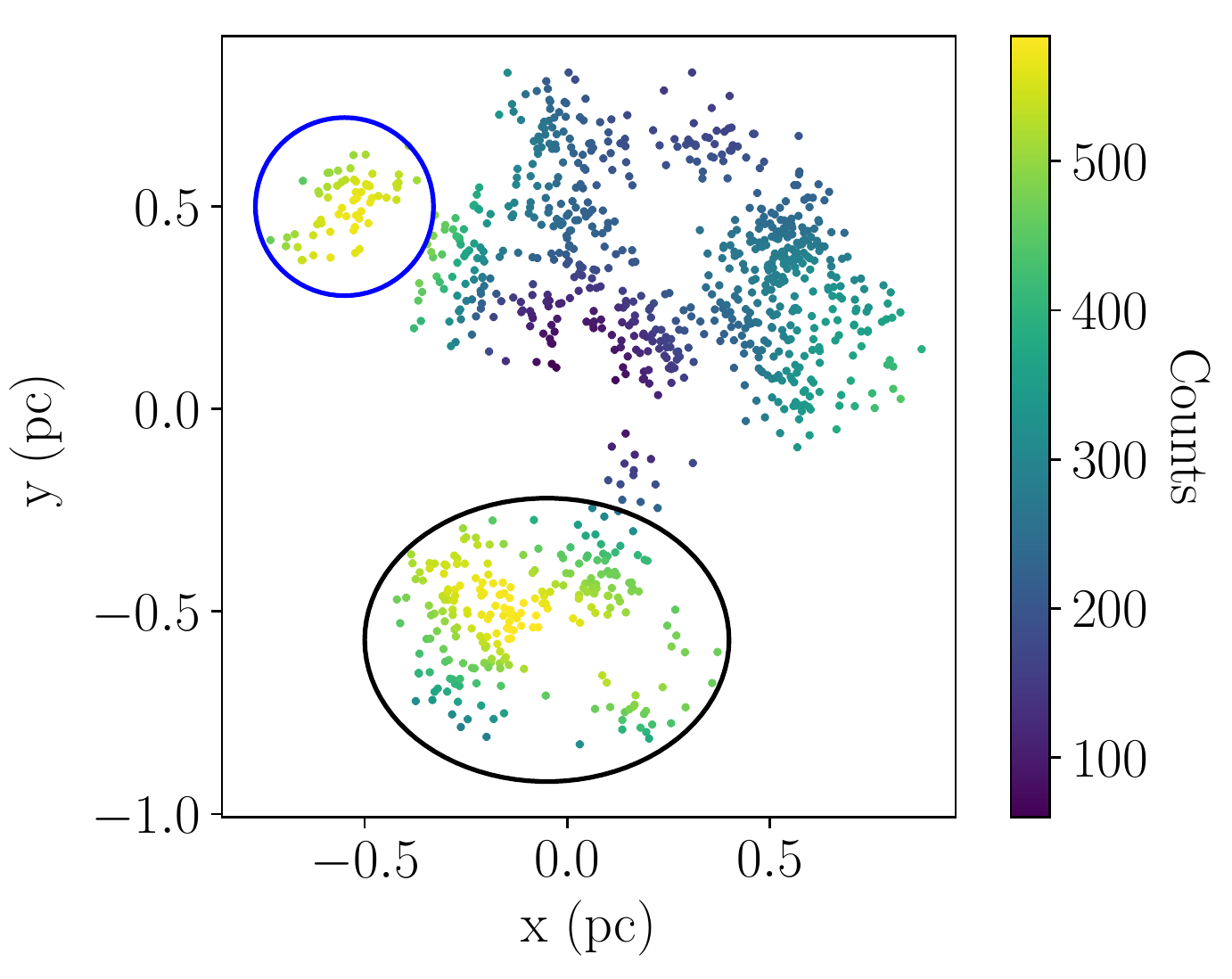}
                 {\phantomsubcaption\label{fig_13_b}}
         \end{subfigure}
         \caption{This figure shows the highly sustructured distribution from
           Fig. \ref{fig_11}. The stars are colour coded
           according to how many times they appear in $\Delta r$ bins between
           0.3 and 0.6 pc (top panel), and between 0.8 and 1.2 pc (bottom panel).}
         \label{fig_13}
\end{figure}

To interpret these features we can consider which stars contribute more than
others in these separation ranges. For example, if a star is in
densely populated area it would
be part of many low $\Delta r$ pairs and would appear many times in
low $\Delta r$ bins. Understanding which stars are contributing most heavily
to the interesting regions of the velocity structure (in this example's case
0.3--0.6 pc and 0.8--1.2 pc) helps us to understand the structure.  Accordingly, the number of times each star appears in $\Delta r$ bins between
0.3 and 0.6 pc is counted. The fractal is plotted with the stars colour coded
by their counts in these bins in the top panel of Fig. \ref{fig_13}. The same is done for the
$\Delta r$ bins between 0.8 and 1.2 pc in the bottom panel of Fig. \ref{fig_13}.
  
First we will look at the simpler case, which for this distribution is the
$\Delta r$ 0.8--1.2 pc range, where $\Delta v_{\rm D}$ is negative.
Inspection of the bottom panel of Fig. \ref{fig_13}
shows that two clumps contribute strongly to these bins. These clumps have
been circled in blue and black on the figure for clarity. 
Comparison of this figure with  Fig. \ref{fig_11}
shows that these clumps are moving 
towards each other, therefore 
$\Delta v_{\rm D}$ is negative in this $\Delta r$ range.  From this analysis we can anticipate that these clumps will
continue to move towards one another (at least in the short term, and in the 2D plane we are observing -- we have no idea here about the third dimension of either position or velocity).
  
The 0.3--0.6 pc range is more complicated. Inspection of the top panel of Fig.
\ref{fig_13} shows that stars in a small clump at
coordinates around (0.15, 0.15) pc which has been circled in black
contribute most often to these bins.
The stars in several surrounding clumps also contribute significantly,
and these clumps have also been circled for clarity.
  
By comparing Fig. \ref{fig_11} and the top panel of Fig.
\ref{fig_13} we see that the stars the in central clump (black circle)
have a bulk motion downwards on the figure (this direction is
defined as `south' for simplicity). To the north there are two clumps,
one circled in orange which is moving to the northwest, and one circled in
blue moving east. Therefore these three clumps are all moving away from
each other, resulting in $\Delta v_{\rm D}$ being positive.
In particular the clump circled in orange is moving directly away from
the main body of the distribution. In the short term we would expect this clump to continue to separate from
the majority of the distribution (at least in this projection).
  
There is one other clump with stars which contribute significantly to
the 0.3--0.6 pc $\Delta r$ bins,
which is in the south and circled in green. This clump is moving
northeast, directly towards the central clump and the clump circled in blue
(so negative $\Delta v_{\rm D}$) and away from the clump circled in orange (positive
$\Delta v_{\rm D}$). Although the $\Delta v_{\rm D}$ contribution
from stars in this clump is mostly negative the number of stars
it contains is small, so it is easy to explain why the mean $\Delta v_{\rm D}$
in the 0.3-0.6 pc $\Delta r$ range is positive. It seems likely that the black
and green circled clumps will continue to move towards each other in the short
term.

In summary with only the raw stellar positions
and velocities shown in Fig. \ref{fig_11} 
the complex velocity structure of the distribution is very difficult to
understand or make judgements on by eye. The method presented in this paper
has been used to explore and interpret the dynamical state of this distribution
and make predictions about its short-term future. 

For the purpose of comparison nine additional highly substructured regions are generated using
the same method but different random number seeds. These region's velocity structures are 
shown in Fig. \ref{fig_14} where the top panel shows
$\Delta v_{\rm M}(\Delta r)$ and the bottom panel $\Delta v_{\rm D}(\Delta r)$.

The first feature of note is that there is much less structure
in both panels of
Fig. \ref{fig_10} than in their corresponding panels in
Fig. \ref{fig_14}, which reflects
the significant velocity substructure in
this latter set of distributions. 
This is useful because while it is easy to distinguish the differing levels of
spatial structure in Fig. \ref{fig_8} and
Fig. \ref{fig_11} by eye the distributions are too
complex to tell simply by looking if the velocity structures
are different. Therefore even if
the actual degree of velocity structure in each set of distributions
were unknown we could still say with
confidence that there is significantly more velocity structure in
this latter set.

We also note that in Fig. \ref{fig_14}
each individual line in both panels appears quite
different from the others. This is unsurprising as
the distributions are produced using different random
number seeds so each is unique, and the
distributions are highly substructured so two statistically
identical distributions may have very different forms \footnote{This raises the question as to if these distributions are indeed `the same', however that is a discussion beyond the remit of this paper.}.
  
In the top panel ($\Delta v_{\rm M}$), the velocity structures show
a general upwards trend; although individual structures show significant
deviation from this (as was mentioned in the discussion of 
Fig. \ref{fig_12}) on the whole $\Delta v_{\rm M}$ correlates positivity
with $\Delta r$. 
This increase of $\Delta v_{\rm M}$ with $\Delta r$ is a result of the
box fractal generation method which produces distributions where 
stars that are near one another have similar velocities and stars that
are far apart have very different velocities.
The magnitude of the features on each line makes it difficult
to say with confidence if there is a  plateau at large $\Delta r$.

In the bottom panel, as is the case in Fig. \ref{fig_10}, some distributions have
predominantly negative $\Delta v_{\rm D}$ and some predominantly positive $\Delta v_{\rm D}$ as
the box fractal method is not biased towards making either expanding or collapsing 
distributions.

\section{Including observational uncertainties} \label{apply_obs_errs}
  
In this section we test whether the method is robust when faced
with imperfect data.

The velocity structure of a simulated
star cluster is measured then observational errors are applied to the data
and the velocity structure is re-calculated. The `true' velocity
structure and `observed' velocity structure are then compared.
A simulation with an unusual spatial and velocity evolution is used to
make this more challenging.

\begin{figure}
  \includegraphics[width=\columnwidth]{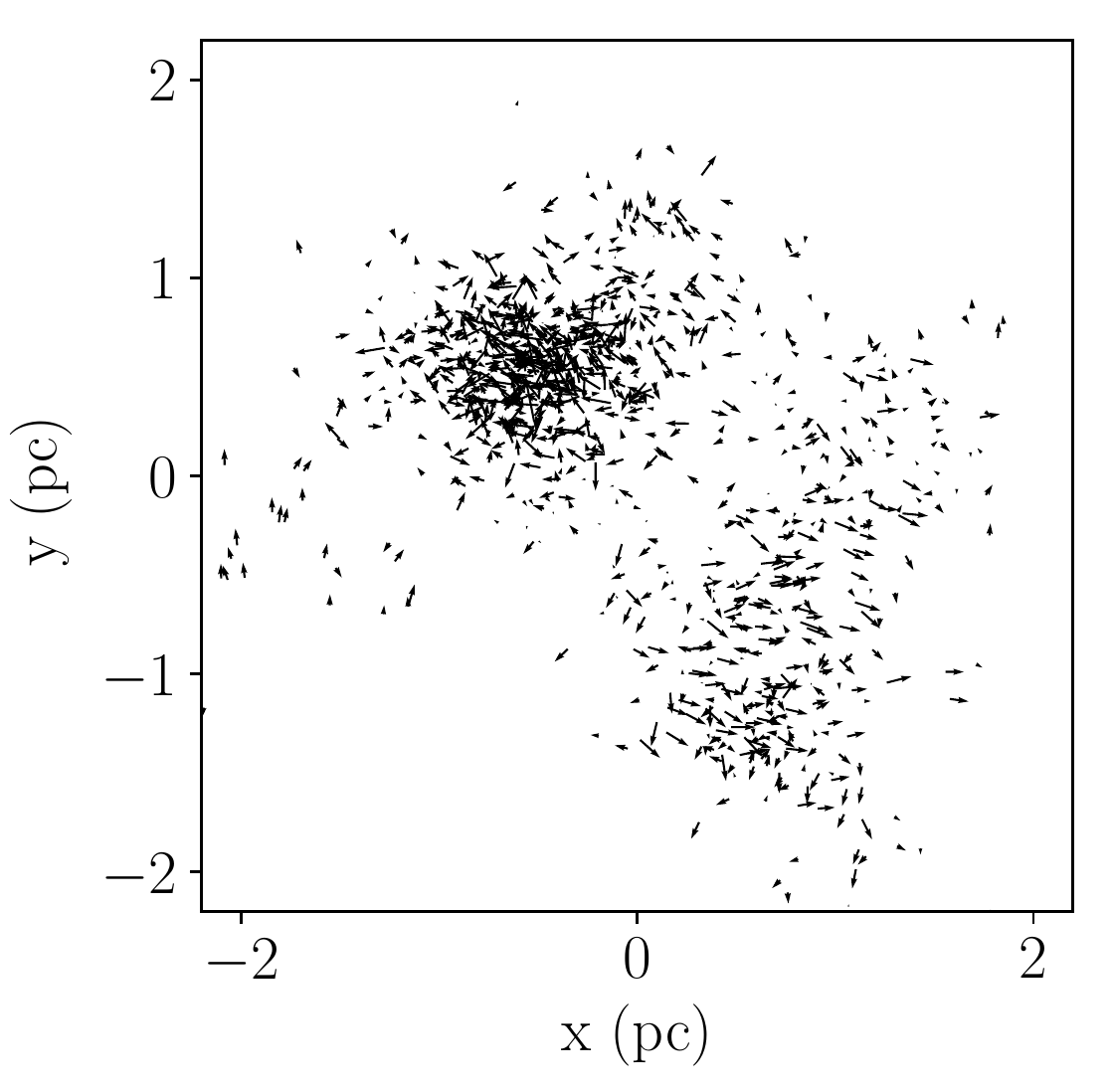}
  \caption{A cluster from a simulation with an unusual velocity evolution.
    Observational errors are applied to this cluster and the `true' and `observed'
  velocity structures are compared.}
  \label{fig_15}
\end{figure}

The cluster is taken from \citet{Arnold17}. That paper gives all 
the details of the simulations, but this cluster contains $N=1000$ 
stars with masses drawn from the Maschberger IMF \citep{Maschberger13} 
using a lower limit of 0.1 M\textsubscript{\(\odot\)} and an upper 
limit of 50 M\textsubscript{\(\odot\)}. It has been evolved for 2 Myr and has split 
into a binary cluster as shown in Fig. \ref{fig_15}. 
  
Although the results presented here concern only this cluster
the same procedure has been applied to a variety of other
simulated clusters, and similar results are found.

\subsection{Velocity uncertainties} \label{sect_obs_uncert}

As stated in section
\ref{Method}, this paper only considers errors on velocities
as they are typically significantly larger than errors on positions.  
We also assume that all stars in the analysis are true members of the cluster. 
Later we remove low-mass stars and examine the impact on the results, but do not add `contaminants' (how important these are will vary significantly depending on the observational dataset).

Observational uncertainties are simulated by replacing each star's
`true' velocity with an `observed' velocity with an associated error. The observed velocity is
drawn from a gaussian centred on the true velocity.
The width of the gaussian used is the observational
uncertainty being simulated, $\sigma_{\rm sim}$  (i.e. the true velocity usually 
lies within the error bar of the observed velocity). 
This is done for the
$x$, $y$, and $z$ components of the velocity separately, i.e. the true $x$ velocity of
a star is replaced with an observed $x$ velocity, etc. Here $\sigma_{\rm sim}$ values 
of 0, 0.4, 0.8, 1.2 and 1.6 km s$^{-1}$ are used.
The $\sigma_{\rm sim} =$ 0 km s$^{-1}$ case is the true velocity
structure as there is no observational uncertainty (although it still has an uncertainty 
associated with stochasticity as in all previous cases).

\begin{figure*}
         \begin{subfigure}[b]{\textwidth}
                 \centering
                 \includegraphics[width=\textwidth]{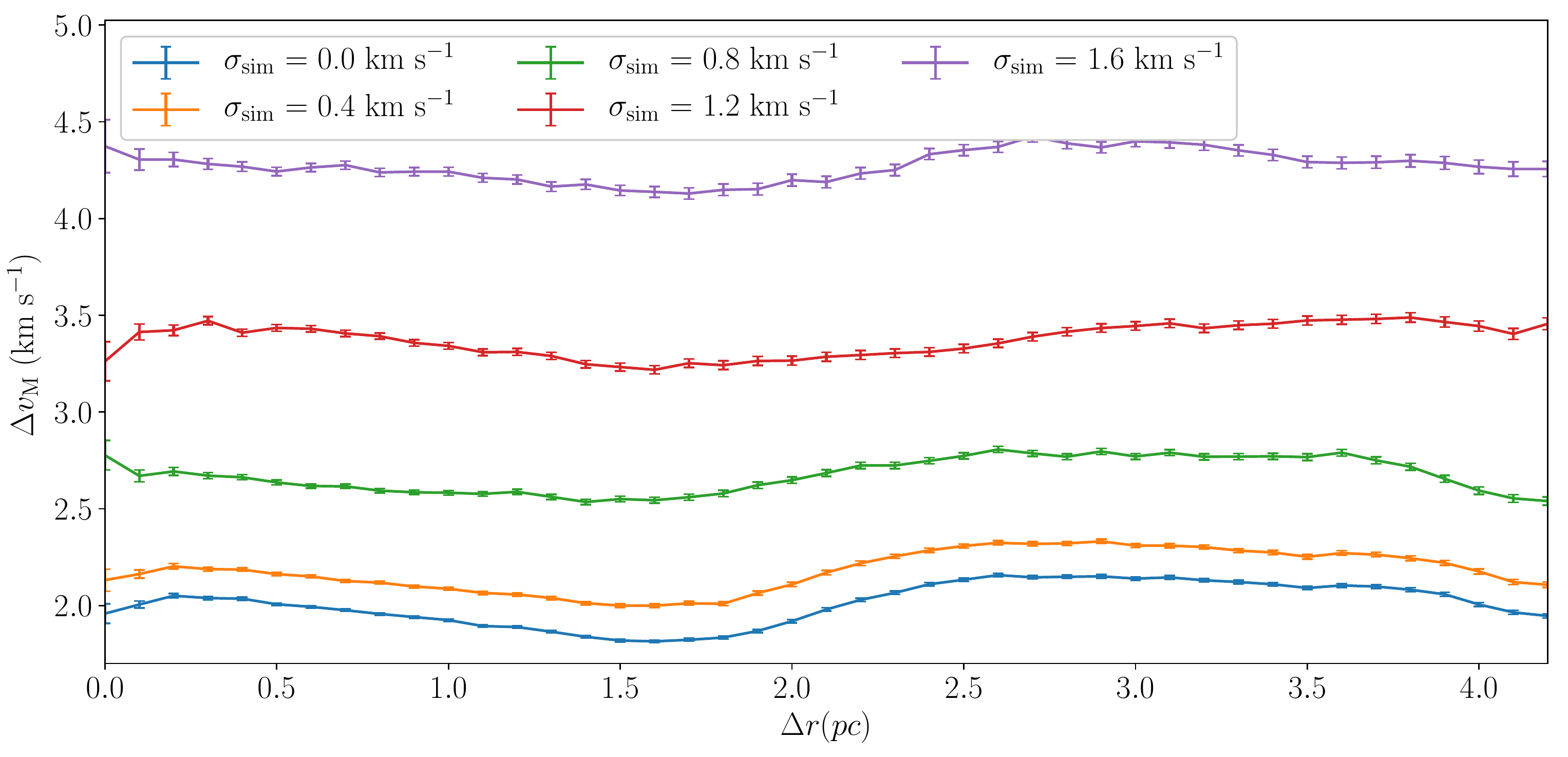}
                 {\phantomsubcaption\label{fig_16_a}}
         \end{subfigure}
         
         \begin{subfigure}[b]{\textwidth}
                 \centering
                 \includegraphics[width=\textwidth]{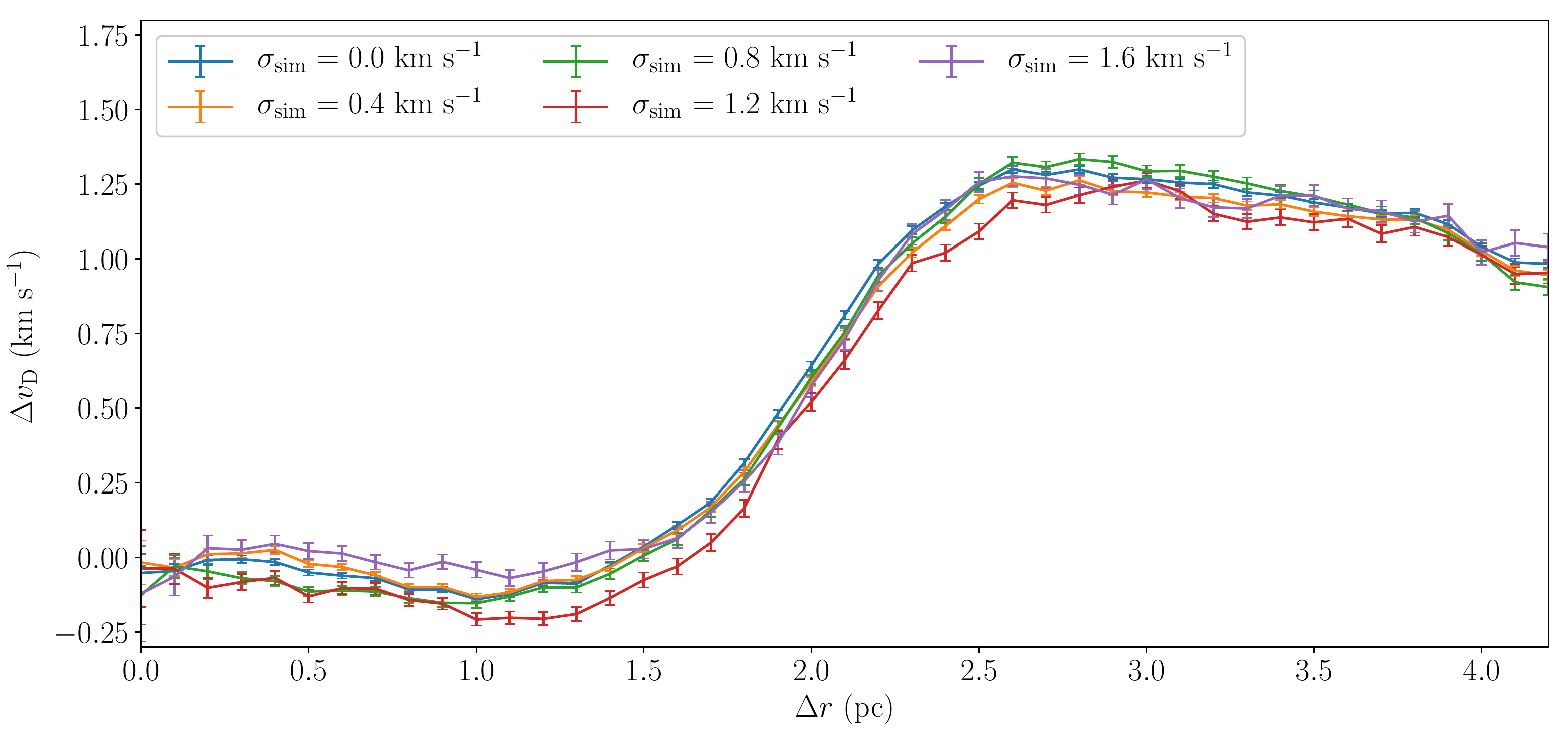}
                 {\phantomsubcaption\label{fig_16_b}}
         \end{subfigure}
         \caption{The velocity structure of the cluster in Fig. \ref{fig_15} with simulated observational
           uncertainties applied. The top panel shows $\Delta v_{\rm M}(\Delta r)$ 
           and the bottom panel shows $\Delta v_{\rm D}(\Delta r)$.
    In both panels a blue line is used for the true velocity structure, orange
    for a simulated observational uncertainty of
    0.4 km s$^{-1}$, green for 0.8 km s$^{-1}$, red for
    1.2 km s$^{-1}$, and purple for 1.6 km s$^{-1}$.}
         \label{fig_16}
\end{figure*}

For each $\sigma_{\rm sim}$ the observed $\Delta v_{\rm M}(\Delta r)$ 
and $\Delta v_{\rm D}(\Delta r)$ is calculated.
These are shown in Fig. \ref{fig_16} where
the velocity structure with $\sigma_{\rm sim}$ = 0  km s$^{-1}$
is shown by the blue line, 0.4 km s$^{-1}$ is the orange line,
0.8 km s$^{-1}$ is the green line,
1.2 km s$^{-1}$ is the red line, and 1.6 km s$^{-1}$ is the purple line.

From inspection of the top panel it is clear that the mean
$\Delta v_{\rm M}$, $\overline{\Delta v_{\rm M}}$,
that is found increases with observational uncertainty from $\sim 2$ km s$^{-1}$ when 
there is no observational error, o $\sim 2.2$ km s$^{-1}$ when the error 
  is $\sigma_{\rm sim}$ = 0.4  km s$^{-1}$,  
  and as $\sigma_{\rm sim}$ increases this trend continues.\footnote{Note that this increase
    is not equal to $\sigma_{\rm sim}\sqrt{2}$ as may be expected.}
The reason for this is that 
uncertainties in the velocities cause the velocity dispersion to be artificially inflated.
As a result the observed difference between any two velocity vectors is more likely to be larger rather than smaller than the `true' difference.

The inflation of $\overline{\Delta v_{\rm M}}$ by observational error
is not of great importance. Much of the useful information
regarding the velocity structure of a cluster
using the magnitude definition is contained in
the shape of the $\Delta v_{\rm M}(\Delta r)$ line, not its placement
on the $\Delta v_{\rm M}$-axis.
Therefore it is reasonable to analyse $\Delta v_{\rm M}(\Delta r)$ 
to investigate a region's velocity structure without
correcting for inflation of $\overline{\Delta v_{\rm M}}$.
Nevertheless, for the interested reader the inflation of 
$\overline{\Delta v_{\rm M}}$ by observational error is discussed
in the appendix, which also describes how this it
can be corrected using Monte Carlo methods.

\begin{figure*}
  \includegraphics[width=\textwidth]{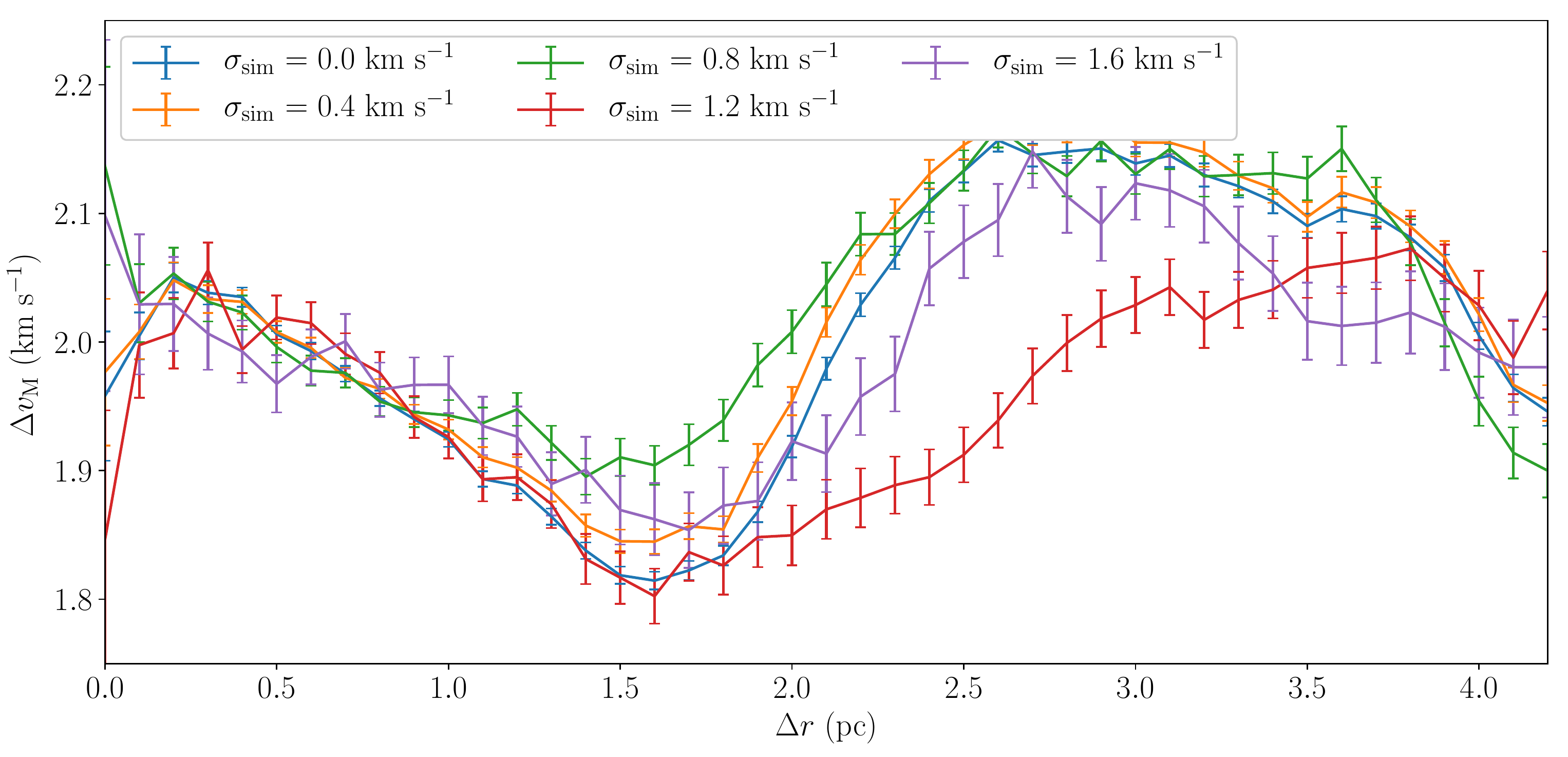}
  \caption{Top panel of Fig. \ref{fig_16} with each line shifted
    such that their $\overline{\Delta v_{\rm M}}$
    matches that of the true velocity structure.}
  \label{fig_17}
\end{figure*}

For the mean time the lines
are shifted such that in every case their $\overline{\Delta v_{\rm M}}$
matches that of the true velocity structure
($\sigma_{\rm sim}$ = 0  km s$^{-1}$)\footnote{The Monte Carlo method works well, but not perfectly. Overlaying the lines exactly allows their features to be compared more easily by eye.}, 
Fig. \ref{fig_17}. 
  
This figure shows a good agreement between
the shape of the observed velocity structures.
As the observational uncertainty increases the
observed velocity structure reproduces
the true velocity structure less well,
but the overall structure remains essentially
recognisable even in the cases where the simulated uncertainty on each
velocity component is greater than the 3D velocity dispersion of
the cluster (1.53 km s$^{-1}$).
From this we conclude that the method deals well with observational uncertainty
up to and potentially beyond the point where the errors are as large as
  the velocity dispersion of the region. 
For $Gaia$ velocity uncertainties depend largely on the apparent magnitude of
  the source. Table B.1 in \citet{Lindegren18} gives median values of these uncertainties
  as a function of apparent magnitude for Gaia DR2.
  For a G-dwarf at $\sim 1$ kpc we would expect errors in proper motion of around 0.3--1 km s$^{-1}$\footnote{The random error in DR2 for G magnitudes of 15--17 is $\sim$0.06--0.2 mas yr$^{-1}$, however there is also a systematic error at the close angular separations we are interested in of $\sim 0.1$ mas yr$^{-1}$ (see \citet{Lindegren18} for details).}.

In the bottom panel of Fig. \ref{fig_16} we show the directional 
velocity structure $\Delta v_{\rm D}(\Delta r)$ (with the lines 
for different errors having the same colours as in the top panel).  
What is obvious here is that the observational errors have essentially 
no effect on the directional structure.  This is because even with uncertainties 
the apparent directions of motion are usually roughly correct, and errors 
between pairs tend to average out rather than sum (as they did above).  
(Note that we assume the errors are uniform across our `field of view', if 
they are not this could introduce a bias but we have not investigated this potential effect.)

\subsection{Mass cutoffs} \label{Mass cutoffs}

A probable bias in observations is to not observe 
low-mass stars as they are typically faint. 
(Note here that larger errors on fainter star's velocities 
would be included in the error propagation). We examine the 
effect of selection limits by removing stars of increasingly 
high mass from our region.

The region has 1000 stars in total which reduces to 428 stars of 
$ >0.3 \: M\textsubscript{\(\odot\)}$, 207 stars of 
$>0.6 \: M\textsubscript{\(\odot\)}$, 128 stars of 
$>0.9 \: M\textsubscript{\(\odot\)}$, and only 83 stars of 
$>1.2 \: M\textsubscript{\(\odot\)}$ (these mass limits are 
rather arbitrary and are just chosen as examples).

\begin{figure*}
         \begin{subfigure}[b]{\textwidth}
                 \centering
                 \includegraphics[width=\textwidth]{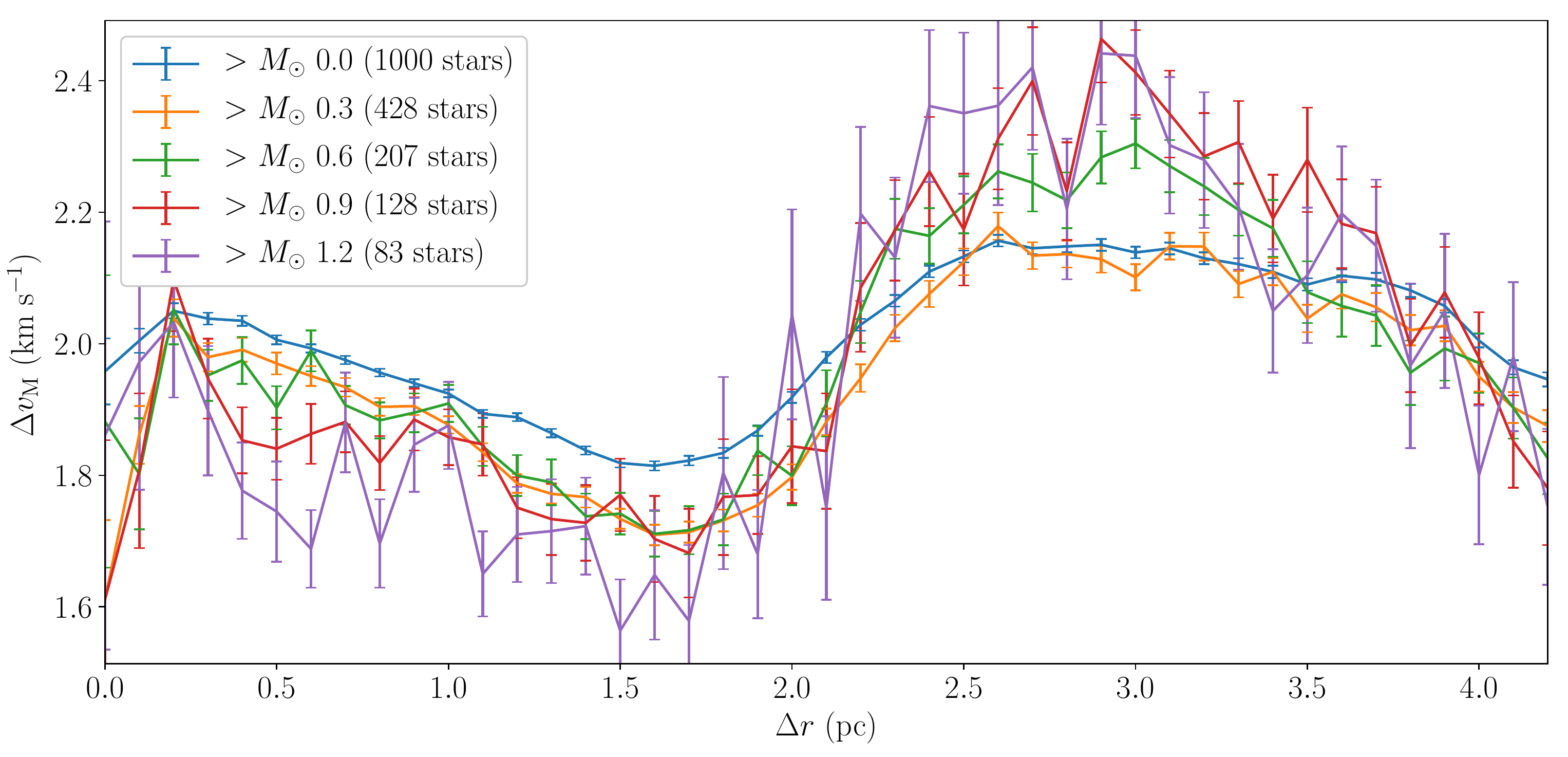}
                 \label{fig_18_a}
         \end{subfigure}
         
         \begin{subfigure}[b]{\textwidth}
                 \centering
                 \includegraphics[width=\textwidth]{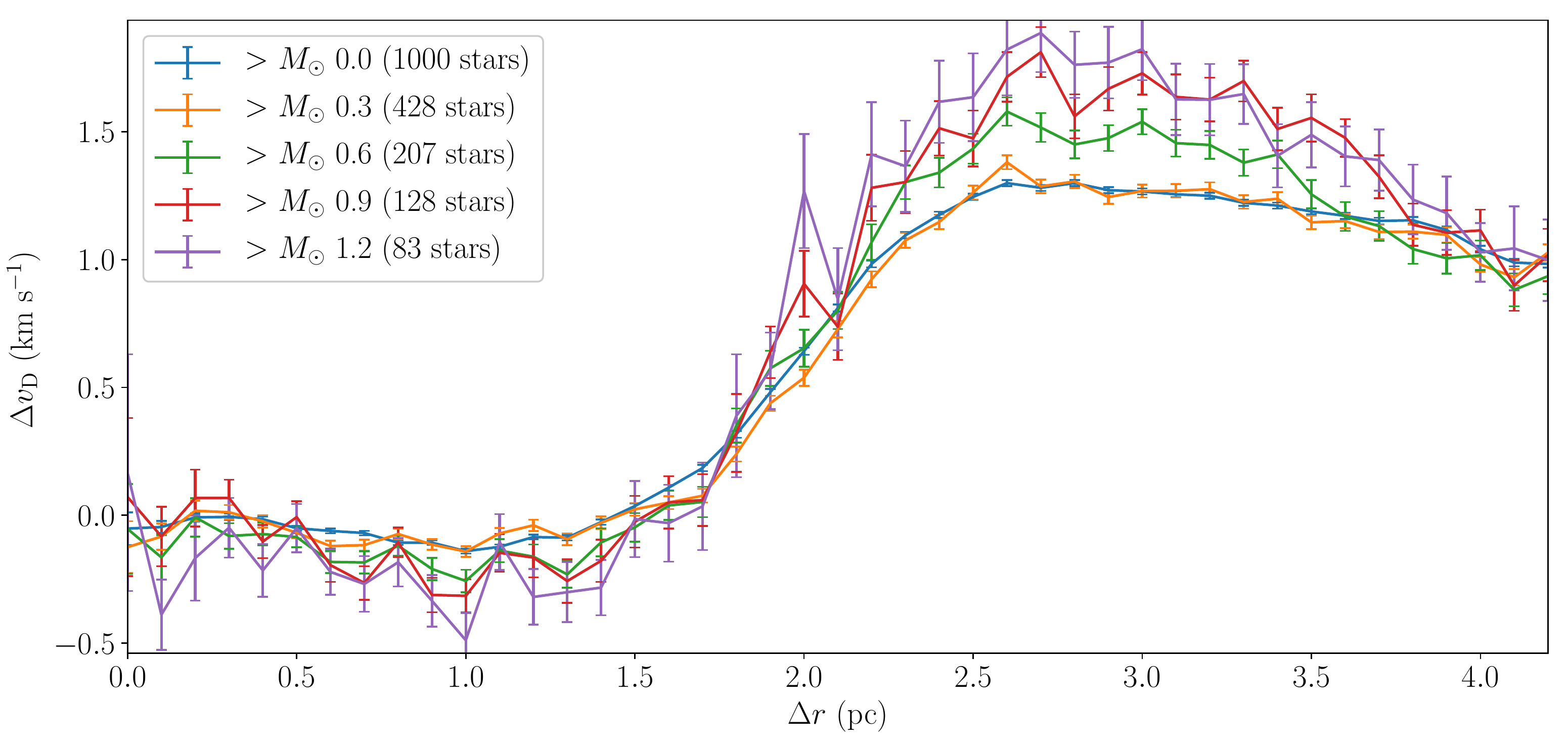}
                 \label{fig_18_b}
         \end{subfigure}
         \caption{The velocity structure of the cluster in Fig. \ref{fig_15} as measured
           by the method using different mass cutoffs. The top panel shows $\Delta v_{\rm M}(\Delta r)$ 
           and the bottom panel shows $\Delta v_{\rm D}(\Delta r)$.
    In both panels the blue line is the result using all stars,
    the orange line using stars above 0.3 M\textsubscript{\(\odot\)},
    green uses those above 0.6 M\textsubscript{\(\odot\)}, red above
    0.9 M\textsubscript{\(\odot\)}, and purple above 1.2
    M\textsubscript{\(\odot\)}.}
         \label{fig_18}
\end{figure*}

Fig. \ref{fig_18} shows the different $\Delta v_{\rm M}(\Delta r)$ (top panel), 
and $\Delta v_{\rm D}(\Delta r)$ (bottom panel) plots.  Different coloured 
lines represent different mass limits as described in the figure.

From Fig. \ref{fig_18} we see that the same basic velocity structure is
observed at all mass limits for both $\Delta v_{\rm M}(\Delta r)$ and
$\Delta v_{\rm D}(\Delta r)$. There does appear to be a sytematic increase
in the amplitude of both $\Delta v_{\rm M}(\Delta r)$ and $\Delta v_{\rm D}(\Delta r)$
at high $\Delta r$ and high mass cutoff. This apparent increase is not observed in other
simulations from the same set. It is therefore determined to be a peculiarity of
this particular simulation like the apparent `kink' observed in Fig. \ref{fig_5}.

The overall robustness of the measured velocity structure against mass cutoffs is encouraging, especially considering
the 1.2 M\textsubscript{\(\odot\)} cutoff leaves only 83 of the cluster's original 
1000 stars remaining, but it is still able to reproduce the shape of the
true underlying velocity structure reasonably well.

That each of our lines for different mass limits are very similar 
shows that in this simulation they all trace a similar velocity `field'. 
This may not be the case in reality, for example in some 
regions the star's spacial and velocity distributions may be a functions of mass
(mass segregated regions being an obvious example).

Nevertheless we can only measure the velocity structure of
the stars which are detected, and from these tests this appears to be robust. 
  
As the mass of the cut-off increases, the level of noise increases 
which is unsurprising as fewer stars survive the higher the cutoff. 
When there are large error bars as a result of low-$N$
the randomisation approach used in section \ref{Applying the method to a Plummer sphere} 
could be used to confirm which features in the observed structure are significant.

\section{Multiple stellar systems} \label{binaries}

So far we have assumed all stars are single.  
However, in observational data and more realistic simulations many stars will
be in binaries or higher-order multiples.
Multiple systems, particularly those in close orbits, often have high
orbital velocities. However, from the point of view of the 
global velocity structure of a region 
a system's centre of mass velocity
better describes the motion of the stars over time than their individual velocities. Here the impact of
binary systems on the velocity structure returned by the method  
is examined (higher order multiples are not included for the sake of simplicity).

This is done by first generating a distribution of 5,000 artificial binary systems.
This large number is chosen to dampen noise due to stochasticity within the distribution. As a 
result fluctuations observed in the results can be confidently attributed to the impact
of binary systems.

The binary systems are generated as follows.
The mass of the primary star is drawn from the
Maschberger IMF \citep{Maschberger13}. The mass ratio of the system is drawn from
a uniform distribution between 0.2 and 1 \citep{Raghavan10} and the mass of the primary
is multiplied by this factor to produce the mass of the secondary. The period, $P$
of the system is drawn from a lognormal distribution centred on log $P$ = 5.03
with a standard deviation of 2.28 (here $P$ is in days) \citep{Raghavan10}. From this the semi-major
axis of the system is calculated. Orbits are circular and the phase and inclination angle
of the system are chosen randomly.  
The position and velocity of the system's centre of mass are also drawn randomly,
the position from a uniform distribution within a 1 pc $\times$ 1 pc $\times$ 1 pc cube,
and the velocity from a gaussian distribution with a standard deviation of 3 km s$^{-1}$
in a random direction.

Synthetic proper motion and radial velocity measurements are then generated from
this distribution. Proper motion measurements are produced by evolving
the distribution forwards by 5 years (gravitational forces exerted on the systems
by each other are neglected because of the shortness of this timescale). The 
change in each star's position in the $x-y$ plane in used to calculate its observed proper motion.
Stars's radial velocities are taken to be their instantaneous velocity in the $z$-direction.

\begin{figure*}
  \includegraphics[width=\textwidth]{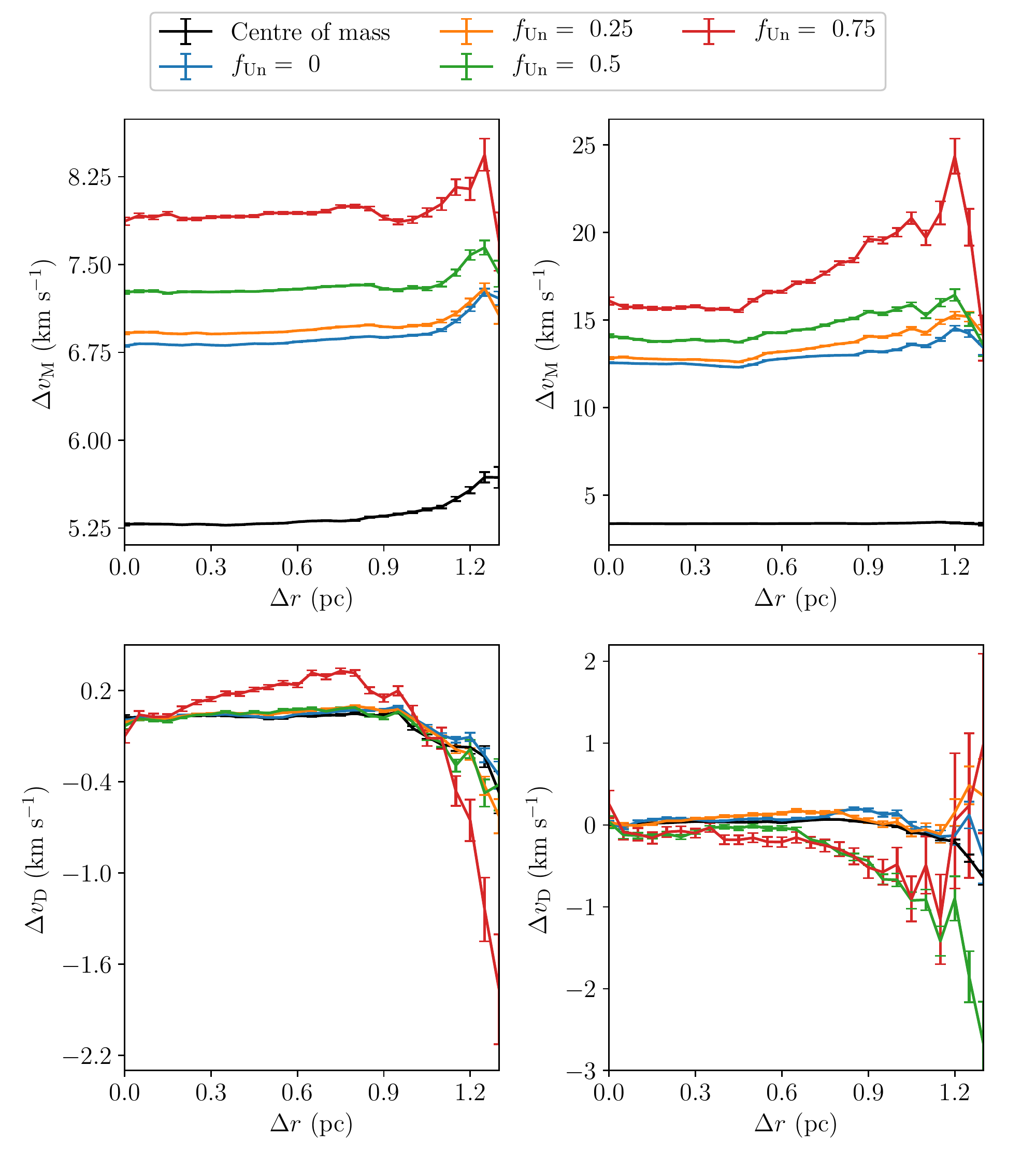}
  \caption{The velocity structure of a distribution of 5,000 binary systems. The top
    two panels use the $\Delta v_{\rm M}$ definition and the bottom two $\Delta v_{\rm D}$.
    The left hand column calculates the velocity structure using synthetic proper motion
    data, and the right hand one synthetic radial velocity data. In all four panels the black
    lines are the velocity structure calculated using the centre of mass velocities of the
    binary systems. The other lines use the velocities of the individual stars.
    The blue lines are the velocity structure calculated when all stars in the sample are observed (the
    unobserved fraction $f_{\rm{Un}}$ is zero). The orange lines are the results when $f_{\rm{Un}}$ is 0.25,
    the green when $f_{\rm{Un}}$ is 0.5, and the red when $f_{\rm{Un}}$ is 0.75.}
  \label{fig_19}
\end{figure*}

In Fig. \ref{fig_19} we show the velocity structure of this distribution of binaries.  
In the left column are the results using proper motion (2D) velocities, and in the right column are results using radial (1D) velocities velocities. The top row uses $\Delta v_{\rm M}$ for each case, and the bottom row uses $\Delta v_{\rm D}$.

In all four panels the results using the system centre of mass velocities are shown by 
black lines. The centre of mass velocities more accurately describe the
  distribution's underlying velocity structure than the velocities of the individual stars
  which contain an orbital component.
All four of these black lines are generally flat as expected for a 
random velocity field.  There is a slight deviation from this at large $\Delta r$ because as $\Delta r$ increases fewer and fewer systems
in the 1 pc square box are sufficiently far apart to populate these bins, making them vulnerable
to stochasticity (see earlier).

The other coloured lines in Fig. \ref{fig_19} are the velocity structure recalculated using the individual velocities of (some) stars.  To model observational limitations we remove some fraction, $f_{\rm{Un}}$, of the lowest-mass (hence lowest luminosity) stars.  For $f_{\rm{Un}}=0.$ (blue lines) all primaries and companions are observed. For an unobserved fraction $f_{\rm{Un}}=0.25$ (orange lines) the 25 per cent lowest-mass stars are `unobserved' and 
are not included in the velocity structure calculation, 
and similarly for $f_{\rm{Un}}=0.5$ (green lines), and $f_{\rm{Un}}=0.75$ (red lines).

Note that (as described earlier) as the size of the region is 1 pc-by-1 pc any features on scales greater than 1 pc should be ignored (or at least taken with extreme caution).

We also note that the results described here reflect the impact of binary stars 
in the worst case scenario:
the binary fraction is 100 \%, and only a single epoch of radial velocity data is used.
Nine other distributions, each with 5,000 binary systems, are produced and analysed as
described here. Their results show the same general trends as the one presented in this paper.

We will first discuss the results using $\Delta v_{\rm M}$ (top row of Fig. 
\ref{fig_19}). In both the case where the proper motions (left panel) and the radial velocities 
(right panel) are used the flat shape of the centre of mass determination of $\Delta v_{\rm M}(\Delta r)$
(the black lines) is largely retained by the results using stellar velocities (coloured lines).
As is to be expected this agreement is poorer when $f_{\rm{Un}}$ is high (and so more stars are 
unobserved), and at large $\Delta r$ (where bins contain fewer pairs and the impact of a 
small number of stars can be more important). As a result artificial structure is visible at high
$f_{\rm{Un}}$ and $\Delta r$. In Fig. \ref{fig_19}, particularly in the radial velocity case, this artificial structure predominantly increases $\Delta v_{\rm M}$. In the nine other realisations
of the distribution, however, there is an even spread between cases where the artificial 
structure increases and decreases $\Delta v_{\rm M}$.

It is clear from the figure that the results using stellar velocities are off-set to 
higher $\Delta v_{\rm M}$. This is due to an inflation of the `velocity dispersion' from the extra
velocity components from binary motion. The degree of the inflation is larger in the radial 
velocity case than the proper motion case as orbital motions, particularly in tight binaries, 
can add significant instantaneous component to the stellar velocity but these are somewhat 
`washed out' by the time baseline of proper motion observations.  
As discussed in section \ref{sect_obs_uncert} the inflation of $\Delta v_{\rm M}$ has minimal
impact on the interpretation of the distribution's velocity structure. 
Overall the agreement between the velocity structure of the 
region as calculated using the centre of mass velocities, and the structure using the stellar 
velocities is good for all but the highest $f_{\rm{Un}}$ and $\Delta r$.

The bottom row of Fig. \ref{fig_19} shows $\Delta v_{\rm D}(\Delta r)$ using proper 
motions (left panel) and radial velocities (right panel).  In both cases the directional velocity
structure is extremely similar for the centres of mass (black lines), and complete or fairly 
complete binary samples (blue and orange lines): a flat distribution at zero $\Delta v_{\rm D}$.
When half, or more, of low-mass stars are unobserved ($f_{\rm{Un}}=0.5$ green 
line, $f_{\rm{Un}}=0.75$ red line) some artificial structure appears.
For most $\Delta r$s this structure has an amplitude below 0.3 km s$^{-1}$, so would almost certainly be lost in the noise of real data.
As in the $\Delta v_{\rm M}$ results the artificial structure can both
increase or decrease $\Delta v_{\rm D}$, and is most severe at high $\Delta r$.

For the case presented in Fig. \ref{fig_19} the $f_{\rm{Un}}=0.5$ results using the proper motions (left panel, green line) is startlingly well behaved. This is a quirk of
the binary distribution presented here, in general there is some artificial structure in the
$f_{\rm{Un}}=0.5$ results. In the radial velocity results there is more deviation, which is more typical.

It is worth reiterating that in the case of radial
velocities a single epoch of observations is assumed. If there were multiple epochs an observer
could potentially estimate binary system's centre of mass velocity, even if only one star is observed.
If an orbital solution cannot not be found but a fluctuation in a star's radial velocity is
observed the suspected binary could be removed from the dataset. This prevents contamination of the calculated velocity
structure by an unknown orbital component and, as was shown in section
\ref{Mass cutoffs}, the method is robust even when a high fraction of stars are not observed.

\begin{figure*}
  \includegraphics[width=\textwidth]{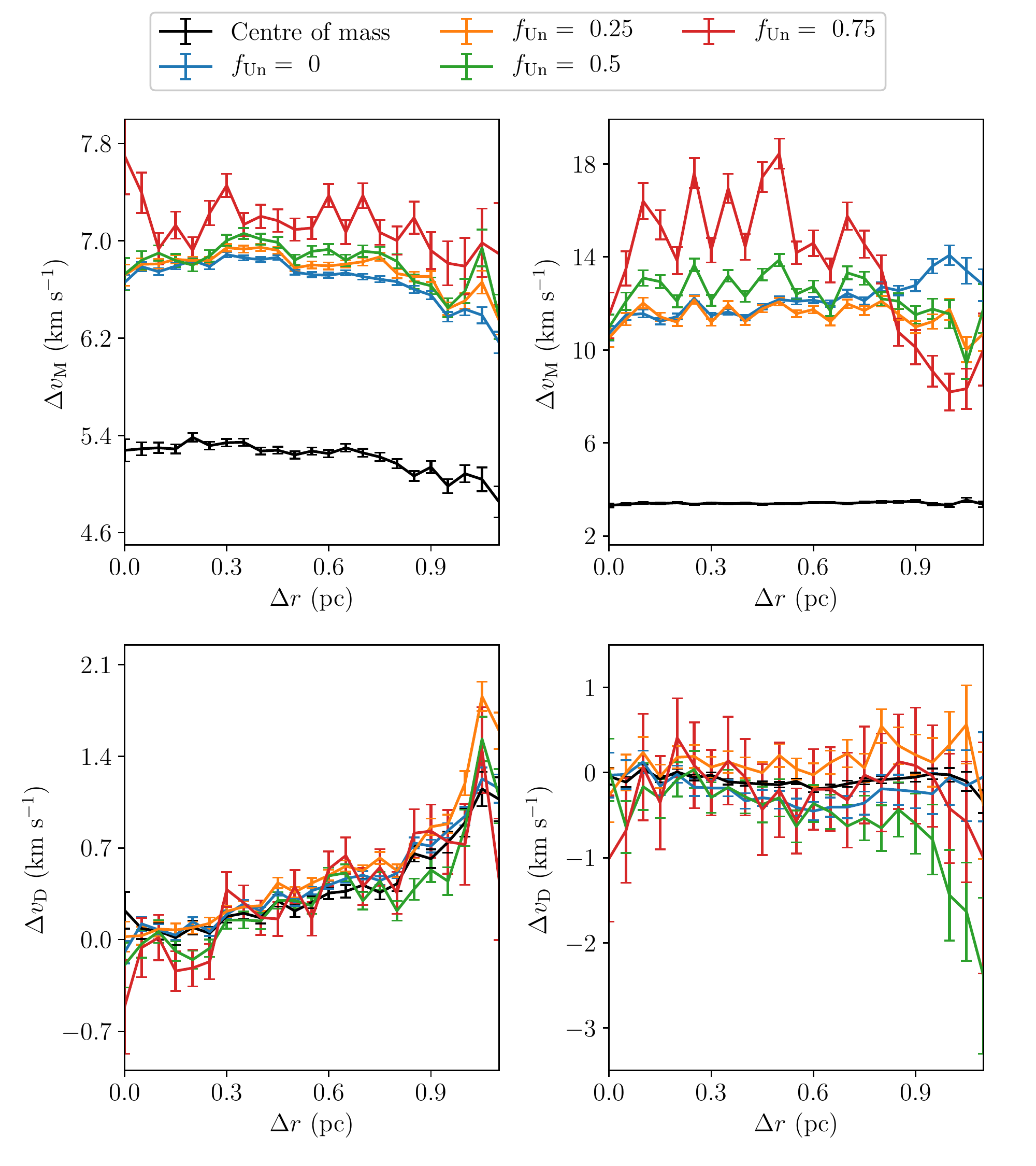}
  \caption{This figure has the same structure as Fig. \ref{fig_19} but it shows
    the velocity structure of a distribution of 500 binary systems rather than 5,000.
    The top panels use the $\Delta v_{\rm M}$ definition and the bottom panels $\Delta v_{\rm D}$.
    The left column calculates the velocity structure using proper motions, and the
    right hand one uses radial velocities.
    The black lines show the velocity structure calculated using the binary system's centre of mass velocities, and the other lines
    are the velocity structure as calculated using the velocities of the individual stars.
    The blue lines use all stars in the sample, the orange lines are the results when $f_{\rm{Un}}$ is 0.25,
    the green when $f_{\rm{Un}}$ is 0.5, and the red when $f_{\rm{Un}}$ is 0.75.}
  \label{fig_20}
\end{figure*}

As described there are 10,000 stars in the distribution used to produce Fig. \ref{fig_19}
  and this large $N$ is chosen to dampen noise due to the stochasticity in the distribution
  (except, as discussed, at high $\Delta r$ where $n_{\rm pairs}$ unavoidably becomes low).
However, many observational datasets have much lower $N$. 
For comparison the procedure described above is repeated for a distribution 
of 1,000 stars (500 binary systems). The results are shown in Fig. \ref{fig_20}.

The velocity structure as calculated using the system's centres of mass velocities is
less flat than in Fig. \ref{fig_19} due to the increase in stochasticity
caused by lower $N$. The velocity structure of the systems
themselves is not of interest here however; it is the degree of agreement
between it and the velocity structure calculated using the stellar velocities
that is being examined.

Inspection of Fig. \ref{fig_20} shows the results
 are noisier and have larger uncertainties than those in
Fig. \ref{fig_19} which can be attributed to the lower $N$.
Nevertheless the agreement is relatively good between the results using centre of mass velocities and stellar velocities although,
as was the case in Fig. \ref{fig_19}, this becomes worse at high $\Delta r$ and $f_{\rm{Un}}$,
and there is an increase in $\Delta v_{\rm M}$ with $f_{\rm{Un}}$.  Again, nine other distributions of 500 binary systems were generated and show the same general trends as Fig. \ref{fig_20}.

We now summarise of the effect of binaries.  Binaries `inflate' $\Delta v_{\rm M}$ with respect to the binary centre of mass determination (exactly by how much depends on the binary population), however the overall structure of $\Delta v_{\rm M}$ remains similar even when a significant fraction of low mass stars are unobserved.  The level and structure of $\Delta v_{\rm D}$ remains very similar, though there are deviations when the `unobservable' fraction is very high.

What is recomforting is that the VSAT method is capable of extracting real structure from even a single epoch of radial velocity data contaminated with binary motions.  Such an analysis should be treated with rather more caution than proper motion data or multi-epoch radial velocity data, but it still contains useful information.

\section{Conclusions} \label{Conclusions}

In this paper we present a method of examining the velocity structure of star
forming regions by plotting the physical separation of pairs of stars ($\Delta r$) 
against their mean velocity difference ($\Delta v$). 
Distributions of $\Delta v(\Delta r)$ for different regions can be directly compared to 
each other. Two definitions of $\Delta v$
are used,  the 'magnitude' definition ($\Delta v_{\rm M}$), and the `directional' definition ($\Delta v_{\rm D}$).

This method does not require the region's centre or radius to be
defined, requires no assumptions about the region's morphology, and can be applied
to data in any number of dimensions in any frame of reference. The method also includes the 
treatment of observational errors, and is shown to be useful even for data with large errors.

The output from the method requires some interpretation, and we have shown a number of examples of how to interpret more complex data.  This is of particular relevance
as we enter this new era of an unprecedented quantity and quality of velocity data.

Although this method was created for the purpose of investigating velocity
structure in star forming regions it is extremely generic; there is no reason 
the data it is applied to must be $r$ and $v$ of stars. This 
makes it a potential tool for investigating very different datasets.

A Python program which runs the method, the Velocity Structure Analysis Tool, \textsc{\small{}vsat}, can be found at 
\hyperlink{https://github.com/r-j-arnold/VSAT}{https://github.com/r-j-arnold/VSAT}.
In the near future we intend to publish a paper demonstrating the
application of this method to observational data (Arnold et al., in preparation).  

\section*{Acknowledgements}

BA acknowledges PhD funding from the University of 
Sheffield. Thanks also to Murali Haran for useful correspondence and to
Liam Grimmett and Gemma Rate for useful discussions.




\bibliographystyle{mnras}
\bibliography{Complete_manuscipt_file}

\section*{Appendix: Correcting inflation}
  
The increase in $\overline{\Delta v_{\rm M}}$ with uncertainty will now be explained in more detail.
As only the magnitude definition of $\Delta v$ is affected 
the M subscript will be dropped to avoid overly long subscripts in this appendix.
  
\begin{figure}
  \includegraphics[width=\columnwidth]{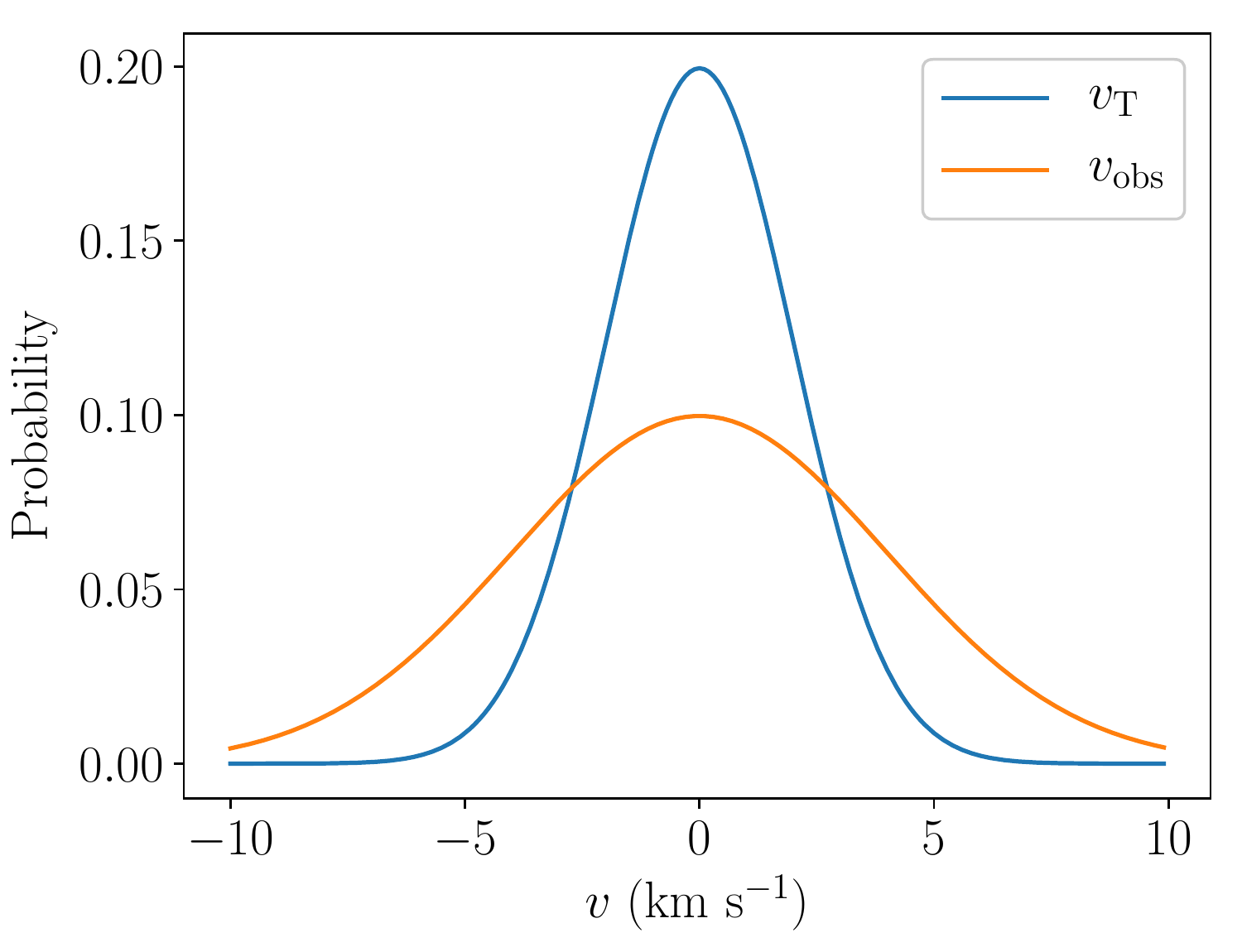}
  \caption{Cartoon depicting the broadening of the observed
    velocity distribution due to observational uncertainties. The x
    axis shows a range of velocities and the y axis their probability. A true
    velocity distribution (in blue) is broadened into the observed velocity
  distribution (in orange).}
  \label{fig_21}
\end{figure}

The true velocities of stars in a region ($v_{\rm T}$)
have some distribution. A cartoon, idealised picture of this
is shown by a blue line in Fig. \ref{fig_21}, where the x axis is
velocity, and the y axis is the probability of a star having a given velocity.
Due to observational uncertainties it is
impossible to perfectly measure the true velocities $v_{\rm T}$, and instead we
observe velocities $v_{\rm obs}$. The effect of observational uncertainties is
to smear out the true velocity distribution. The observed velocity distribution is
shown by the orange line
in Fig. \ref{fig_21} for our cartoon case. Notice that the
observed velocity distribution is wider that the
true velocity distribution.
  
When the velocity difference between two stars in a region is measured this
can be thought of as drawing two velocities from the velocity
distribution and calculating the difference between them.
If the distribution is narrow then the range of
likely velocities is small so the two velocities
drawn will usually have a small difference between them, therefore
$\overline{\Delta v}$ will be small. In contrast, if the distribution is
wide it is more likely that any two values drawn will be very different, so
$\overline{\Delta v}$ will be large. As discussed, the observed
velocity distribution is wider than the true velocity distribution,
so the observed mean velocity difference between pairs of
stars ($\overline{\Delta v_{\rm obs}}$) is
larger than the true mean velocity difference between pairs of
stars ($\overline{\Delta v_{\rm T}}$).
Because the width of the $v_{\rm obs}$ distribution increases
with uncertainty so does $\overline{\Delta v_{\rm obs}}$. This is why in the top panel
of Fig. \ref{fig_16} there is a positive correlation between $\overline{\Delta v}$
and $\sigma_{\rm sim}$.
  
As discussed above observational errors broaden the observed
velocity distribution, so the true velocity distribution
can be crudely approximated by a narrower
version of the observed velocity distribution. In brief, the observed
velocity distribution is narrowed by different amounts and Monte Carlo
methods are used to find which width best reproduces the observed velocity
distribution once observational errors are applied. Many velocities are then drawn from this
best fitting distribution, and $\overline{\Delta v}$ is calculated.
This is the estimated value of $\overline{\Delta v_{\rm T}}$ given
the observed velocities and the errors.
  
\begin{figure*}
    \begin{subfigure}[b]{0.48\textwidth}
    \centering
    \includegraphics[width=1.\textwidth, trim={0 0 0 1cm},clip]{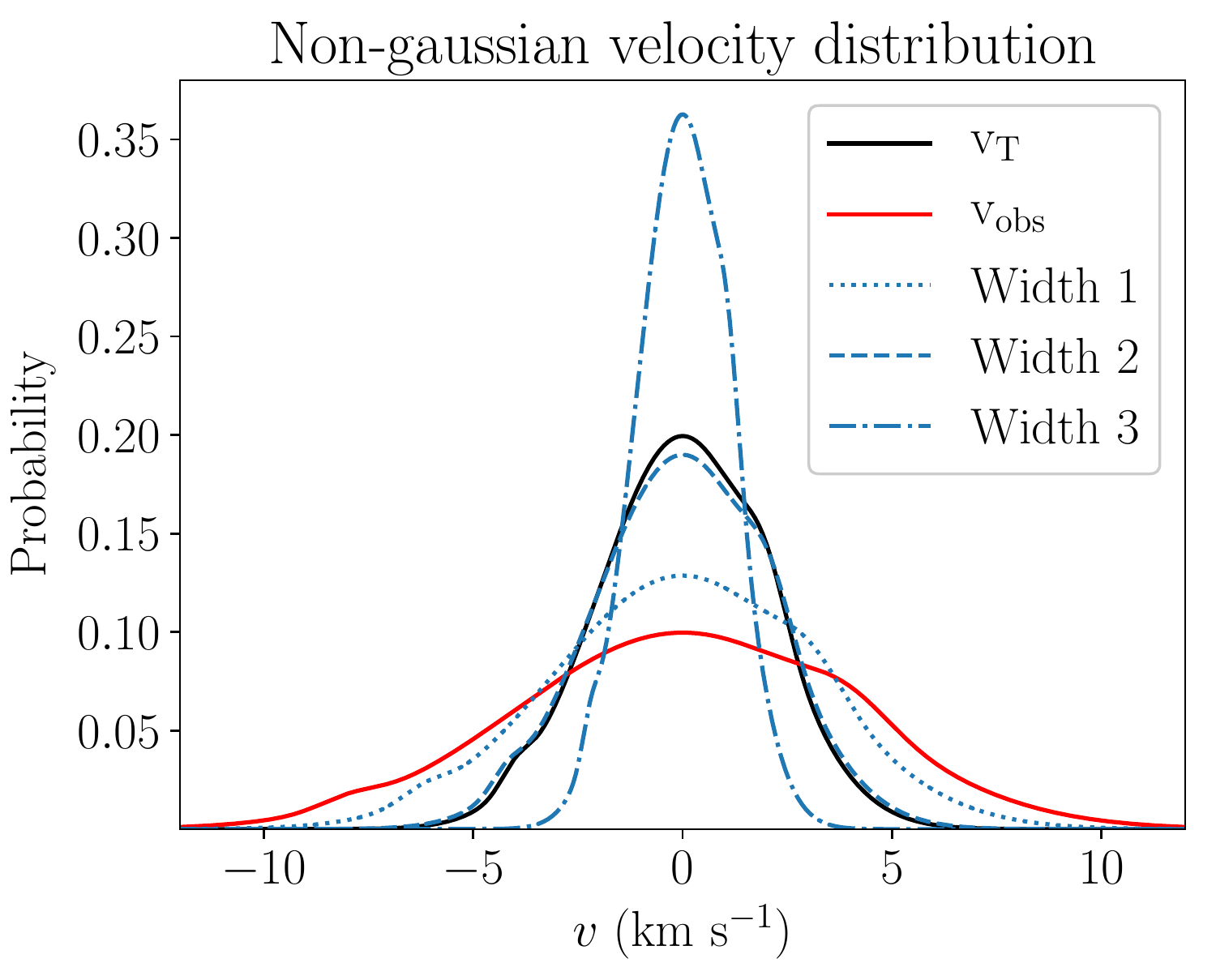}{\phantomsubcaption\label{fig_13_a}}
    {\phantomsubcaption\label{fig_22_a}}
  \end{subfigure}
  \begin{subfigure}[b]{0.48\textwidth}
    \centering
    \includegraphics[width=1.\textwidth, trim={0 0 0 1cm},clip]{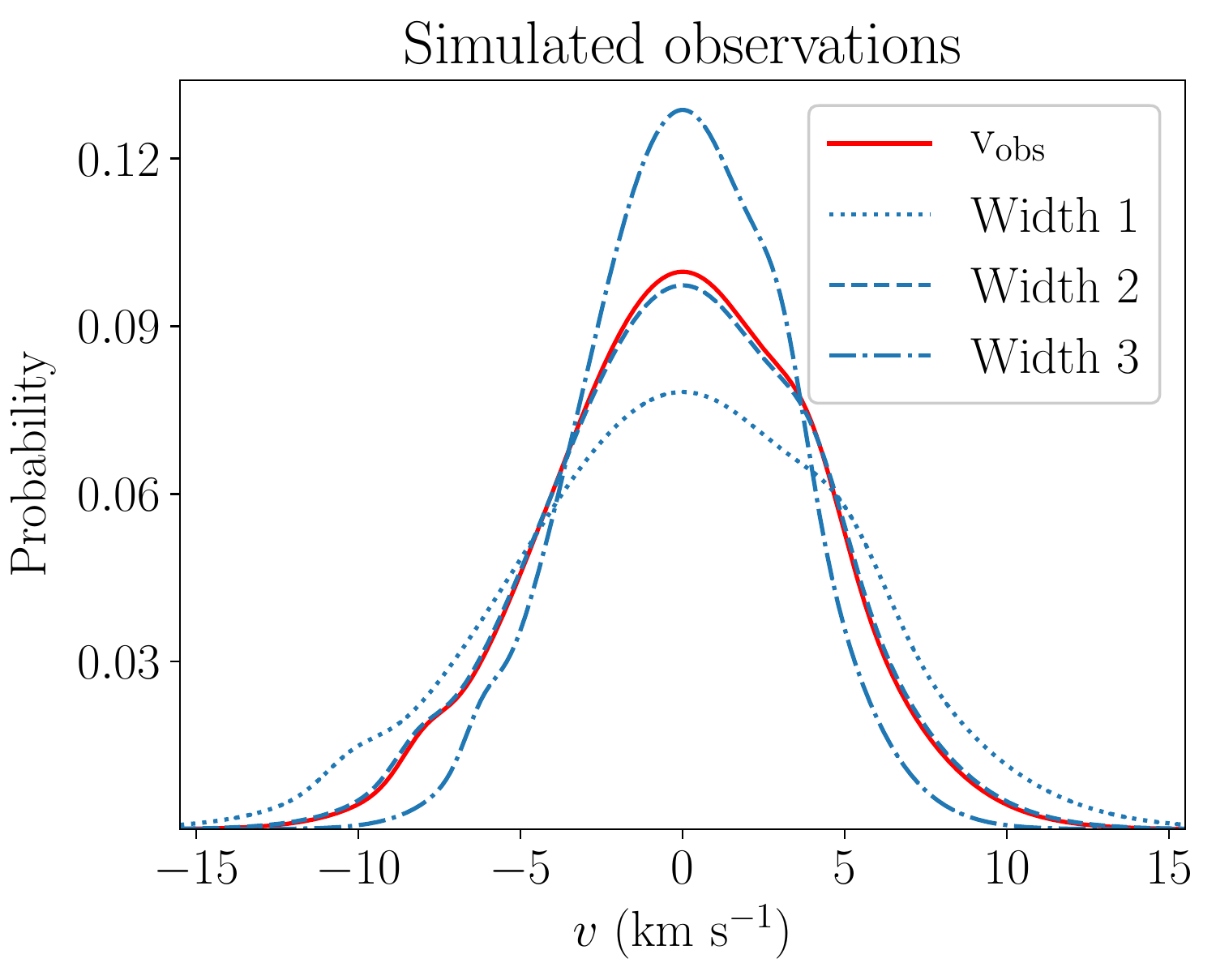}
    {\phantomsubcaption\label{fig_22_b}}
  \end{subfigure}
  \caption{Diagrams aiding the explanation of how to correct for
    $\overline{\Delta v}$ inflation by observational uncertainties.
    For both the left and right panel the x axis is velocity and the y axis
    is probability. The left hand panel depicts a true velocity distribution (black line),
    the observed velocity distribution (red line) and 3 models of the true non-gaussian velocity distribution
    using different widths (blue dashed lines). The right hand panel shows the observed velocity distribution (red line), and the simulated observations assuming each
    of the models from the left hand panel (blue dashed lines).}
    \label{fig_22}
\end{figure*}

The exact method used will now be described in more detail.
Diagrams shown in Fig. \ref{fig_22} are referred
to to aid this description. For both of these plots the x axis is
velocity, and the y axis is probability.
They show how the method would be applied
to some cartoon non-gaussian velocity distribution
(the black line in the left panel of Fig. \ref{fig_22}).  
  
First a gaussian kernel is applied to the observed velocities to produce
a probability density function (pdf) of the observed velocities
(red line in both panels of Fig. \ref{fig_22}). 
It is assumed that the true velocities pdf is the
same shape, but narrower. How much narrower is unknown, and though
it can be analytically calculated if the distributions are gaussian that will
often not be the case. Instead many different widths are tested, each
model being a `guess' at the true velocity structure.
To prevent Fig. \ref{fig_22} becoming overcrowded only three
models are shown (blue dashed lines). In this diagram it is obvious that the first is
much wider than the $v_{\rm T}$ distribution,
the second is almost exactly right, and the third is much narrower.
In reality $v_{\rm T}$ would be unknown, so it is not
so easy to compare.

For each model $N$ velocities are drawn and observational
uncertainties are applied as per the method described earlier
in this section. The distributions of these simulated
velocity observations are what we would expect to observe if
the model were the true distribution.
This is repeated many times (100 in this paper)
in order to obtain reliable results.
The right hand panel of Fig. \ref{fig_22} shows how these simulated observational distributions compare to
the actual observed distribution. If the model the
velocities are drawn from is a good match for the
true velocity distribution then the simulated
observations  distribution will replicate the actually observed distribution well.
From the left hand panel of Fig. \ref{fig_22} it is evident that width 1 is too large,
width 2 is approximately correct, and width 3 too narrow, and
this is reflected in the right hand panel. Clearly the simulated observations using
width 2 is the best match to the observations, and so is
taken to be a good approximation of the true velocity structure.
  
Now the true velocity distribution has been modelled a
large number of velocities are drawn from it and $\overline{\Delta v}$
is calculated. This $\overline{\Delta v}$ is the estimated value
of $\overline{\Delta v_{\rm T}}$.

\begin{figure}
  \includegraphics[width=\columnwidth, trim={0 0 0 1cm},clip]{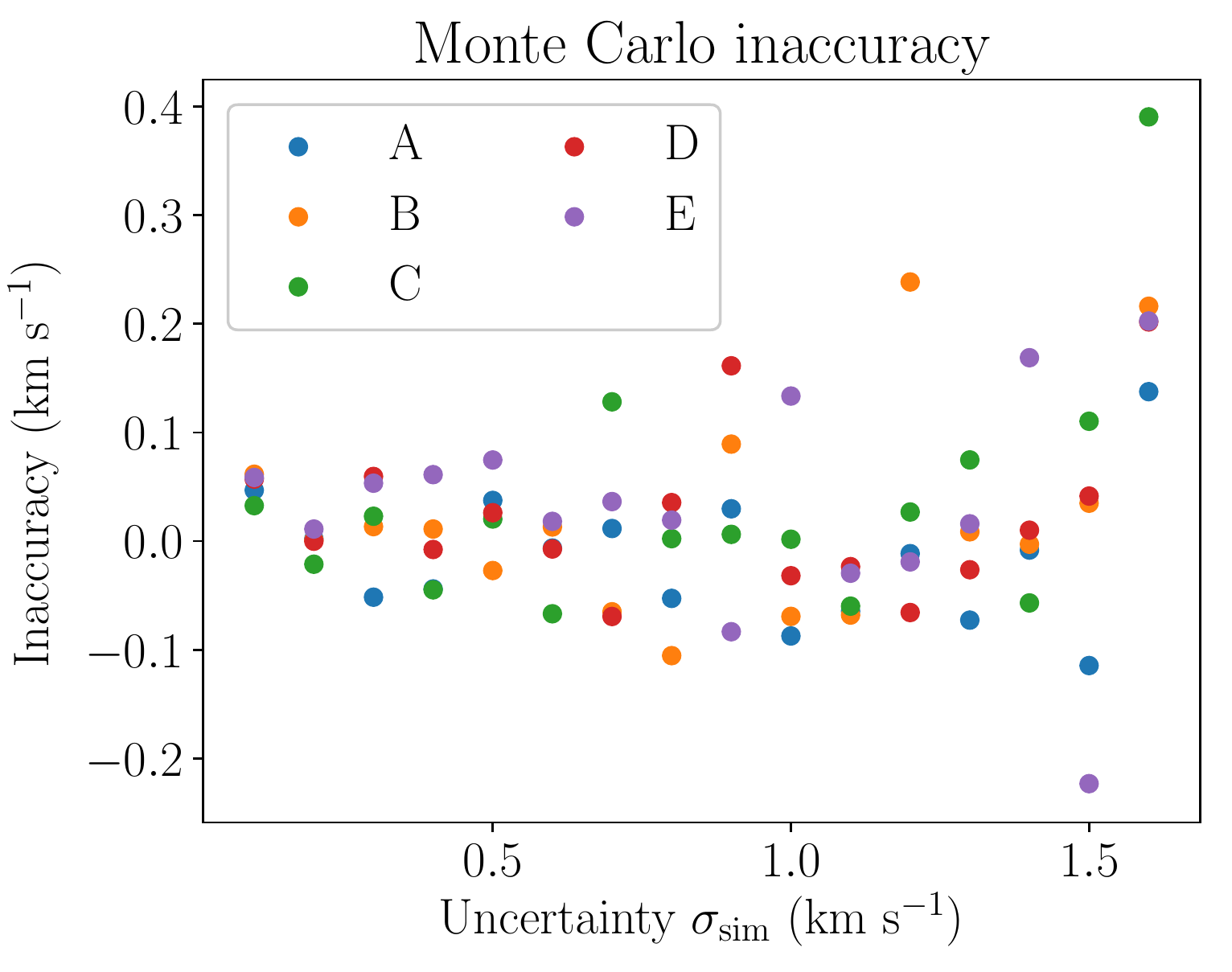}
  \caption{Plot showing the inaccuracy of the value of
    $\overline{\Delta v_{\rm T}}$ estimated using the Monte Carlo method. The x axis
    is the observational uncertainty applied to the data and the y axis is the inaccuracy.
  Different colours are used for each of the five simulations tested.}
  \label{fig_23}
\end{figure}

To quantify how accurate this the method is it is applied
to five very different simulated regions, A, B, C, D, and E.
For each the true $\overline{\Delta v_{\rm T}}$ is calculated, then
observational uncertainties are applied and the Monte Carlo method is
used to estimate $\overline{\Delta v_{\rm T}}$ from the observed velocities.
This is done for observational uncertainties ($\sigma_{\rm sim}$) between 0.1 and 1.6 km s$^{-1}$
in steps of 0.1 km s$^{-1}$. In each case the difference between the true
$\overline{\Delta v_{\rm T}}$ and
the value of $\overline{\Delta v_{\rm T}}$ estimated using the Monte Carlo method
is computed. This difference is referred to as the inaccuracy. For
each of the five simulations inaccuracy is plotted against $\sigma_{\rm sim}$,
which is shown in Fig. \ref{fig_23}.
  
From Fig. \ref{fig_23} we see a rough correlation between $\sigma_{\rm sim}$ and
inaccuracy, which is expected.
More importantly we see that the inaccuracy observed is low, typically $\lesssim$ 0.1
km s$^{-1}$ except for extremely high uncertainties. We therefore conclude that
$\overline{\Delta v_{\rm T}}$ can be recovered from the observed velocities
with reasonably high accuracy. Unfortunately exact error limits can't be
calculated because error is introduced by the assumption that the
true velocity distribution has the exact same shape as the observed
velocity distribution, it is only narrower. This assumption will never be
perfectly true but only close, and without knowing the true velocity structure
it is impossible to know how close. Therefore the error can't be quantified.

Nevertheless it has been shown this method can reproduce
$\overline{\Delta v_{\rm T}}$ with reasonable accuracy if the errors
on the velocity measurements are not too high. Also, as stated earlier,
$\overline{\Delta v_{\rm T}}$ is largely irrelevant to interpretation
of the velocity structure when $\Delta v_{\rm M}$, it is the shape which contains the majority of
the information.

  
  \bsp	
  \label{lastpage}
  \end{document}